\documentclass[final,times]{elsarticle}

\usepackage{natbib}
\biboptions{sort&compress}
\usepackage{graphicx}
\usepackage{subcaption}
\usepackage{amstext}
\usepackage{amssymb}
\usepackage{amsmath}
\usepackage{color}
\usepackage{booktabs}
\usepackage{todonotes}
\usepackage{siunitx}
\usepackage{mhchem}
\usepackage[hidelinks=true]{hyperref}
\usepackage[displaymath]{lineno}

\graphicspath{{./gfx/}}

\renewcommand{\vec}[1]{\ensuremath{\mathbf{#1}}}
\newcommand{\mat}[1]{\ensuremath{\mathbf{#1}}}

\newcommand{\w}{\ensuremath{\text{w}}}
\newcommand{\n}{\ensuremath{\text{n}}}

\newcommand{\advective}{\ensuremath{\text{a}}}
\newcommand{\capillary}{\ensuremath{\text{c}}}
\newcommand{\ca}{\ensuremath{\text{Ca}}}

\newcommand{\dd}{\ensuremath{\text{d}}}
\newcommand{\pd}[2]{\ensuremath{\frac{\partial #1}{\partial #2}}}
\newcommand{\od}[2]{\ensuremath{\frac{\dd #1}{\dd #2}}}

\journal{Frontiers in Physics}

\begin{document}


\begin{frontmatter}

\title{Stable and efficient time integration of a dynamic pore network
  model for two-phase flow in porous media}

\author[ntnu-ify]{Magnus Aa. Gjennestad\corref{mag-corresp}}
\ead{magnus@aashammer.net}
\author[ntnu-ify]{Morten Vassvik}
\author[ntnu-ikj]{Signe Kjelstrup}
\author[ntnu-ify]{Alex Hansen}

\cortext[mag-corresp]{Corresponding author.}

\address[ntnu-ify]{PoreLab and Department of Physics, Norwegian
  University of Science and Technology, \\ Trondheim, Norway}
\address[ntnu-ikj]{PoreLab and Department of Chemistry, Norwegian
  University of Science and Technology, \\ Trondheim, Norway}

\begin{abstract}
We study three different time integration methods for a dynamic pore
network model for immiscible two-phase flow in porous
media. Considered are two explicit methods, the forward Euler and
midpoint methods, and a new semi-implicit method developed herein. The
explicit methods are known to suffer from numerical instabilities at
low capillary numbers. A new time-step criterion is suggested in order
to stabilize them. Numerical experiments, including a Haines jump
case, are performed and these demonstrate that stabilization is
achieved. Further, the results from the Haines jump case are consistent
with experimental observations. A performance analysis reveals that
the semi-implicit method is able to perform stable simulations with
much less computational effort than the explicit methods at low
capillary numbers. The relative benefit of using the semi-implicit
method increases with decreasing capillary number $\ca$, and at $\ca
\sim \SI{e-8}{}$ the computational time needed is reduced by three
orders of magnitude. This increased efficiency enables simulations in
the low-capillary number regime that are unfeasible with explicit
methods and the range of capillary numbers for which the pore network
model is a tractable modeling alternative is thus greatly extended by
the semi-implicit method.

\end{abstract}

\begin{keyword}
porous media \sep two-phase flow \sep pore network model \sep
numerical methods \sep time integration \sep stability \sep low
capillary number

\vspace{0.2cm}
\noindent
\href{https://doi.org/10.3389/fphy.2018.00056}{\textit{doi:} 10.3389/fphy.2018.00056}

\end{keyword}

\end{frontmatter}

\section{Introduction}

Different modeling approaches have been applied in order to increase
understanding of immicsible two-phase flow in porous media. On the
pore scale, direct numerical simulation approaches using e.g.\ the
volume of fluid method \cite{Raeini2012} or the level-set method
\cite{Jettestuen2013,Gjennestad2015} to keep track of the fluid
interface locations, have been used. The lattice-Boltzmann method is
another popular choice, see e.g.\ \cite{Ramstad2010}. These methods
can provide detailed information on the flow in each pore. They are,
however, computationally intensive and this restricts their use to
relatively small systems.

Pore network models have proven to be useful in order to reduce the
computational cost \cite{Hammond2012}, or enable the study of larger
systems, while still retaining some pore-level detail. In these
models, the pore space is partitioned into volume elements that are
typically the size of a single pore or throat. The average flow
properties in these elements are then considered, without taking into
account the variation in flow properties within each element.

Pore network models are typically classified as either quasi-static or
dynamic. The quasi-static models are intended for situations where
flow rates are low, and viscous pressure drops are neglected on the
grounds that capillary forces are assumed to dominate at all times. In
the quasi-static models by \citet{Lenormand1988},
\citet{Wilkinson1983} and \citet{Blunt1998}, the displacement of one
fluid by the other proceeds by the filling of one pore at the time,
and the sequence of pore filling is determined by the capillary entry
pressure alone.

The dynamic models, on the other hand, account for the viscous
pressure drops and thus capture the interaction between viscous and
capillary forces. As three examples of such models, we mention those
by \citet{Hammond2012}, \citet{Joekar-Nisar2010} and
\citet{Aker1998}. A thorough review of dynamic pore network models was
performed by \citet{Joekar-Niasar2012}.

The pore network model we consider here is of the dynamic type that
was first presented by \citet{Aker1998}. Since the first model was
introduced, it has been improved upon several times. Notably, it was
extended to include film and corner flow by \citet{Tora2012}. The
model considered here does not contain this extension. This class of
models, which we call the Aker-type models, is different from the
majority of other pore network models
\cite{Hammond2012,Joekar-Nisar2010} in that both the pore body and
pore throat volumes are assigned to the links, and no volume is
assigned to the nodes. Fluid interface locations are tracked
explicitly as they move continuously through the pore space. This is
in contrast to the model by \citet{Hammond2012}, where interfaces are
moved through whole volume elements at each time step, and to the
model of \citet{Joekar-Nisar2010}, where interface locations are only
implicitly available through the volume element saturation. One of the
advantages of the Aker-type model is that a detailed picture of the
fluid configuration is provided at any time during a simulation.
Dynamic phenomena, such as the retraction of the invasion front after
a Haines jump \cite{Haines1930,Berg2013,Armstrong2013,Maloy1992}, are
thus easily resolved.

Since 1985, numerical instabilities at low capillary numbers have been
known to occur for various types of dynamic pore network models
\cite{Koplik1985}. A whole section is devoted to the topic in the
review by \citet{Joekar-Niasar2012}. It is important to address such
stability problems rigorously, as many of the practical applications
of two-phase porous media flow are in the low capillary number
regime. Examples include most parts of the reservoir rock during
\ce{CO2} sequestration, flow of liquid water in fuel cell gas
diffusion layers and studies of Haines jump dynamics, see
e.g.~\cite{Armstrong2013}.

When Aker-type pore network models are used, the numerical
instabilities are observed as oscillations in the positions of the fluid
interfaces. Some efforts to avoid these oscillations have been made by
introduction of modifications to the model. \citet{Medici2010} used a
scheme where water was allowed to flow in the forward direction only
in order to study water invasion in fuel cell gas diffusion
layers. While this approach led to interesting results, it has some
downsides. First, the interface movement is artificially restricted,
and certain dynamic effects can not be resolved. This includes
e.g.\ invasion front retraction after a Haines jump. Second, the
method can only be used in cases with transient invasion. Studies of
steady-state flow, such as those performed by \citet{Knudsen2002}
and \citet{Savani2017}, are not possible.

Because the oscillations originate in the numerical methods, rigorous
attempts to remove them should focus on these methods rather than the
models themselves. \citet{Joekar-Nisar2010} followed this avenue and
achieved stabilization using a linearized semi-implicit method. Their
work, however, concerned a different type of pore network model than
that considered here.

In this work, we present three numerical methods that can be utilized
to perform stable simulations of two-phase flow in porous media with
pore network models of the Aker type. The stability problems
previously observed are thus solved without the need to resort to
model modifications that restrict interface movement or preclude
steady-state flow simulations. Two explicit methods are discussed, the
forward Euler method and the midpoint method. These are stabilized by
a new time step criterion derived herein. The third method is a new
semi-implicit method. Thorough verifications of all methods are
performed, confirming correct convergence properties and
stability. Finally, we compare the methods in terms of performance.

The rest of this paper is structured as
follows. Section~\ref{sec:model} contains background information on
the pore network model. Section~\ref{sec:methods} presents briefly the
nomenclature, used in subsequent sections to describe the time
integration methods. In Section~\ref{sec:fem}, we recapitulate how the
forward Euler method is used to integrate the model and we present a
new time step criterion that stabilizes both forward Euler and the
midpoint method at low capillary numbers. We briefly review the
midpoint method in Section~\ref{sec:mpm}. The new semi-implicit method
is described in detail in Section~\ref{sec:sim}. Some remarks about
the numerical implementation are made in
Section~\ref{sec:implementation}. Section~\ref{sec:cases} contains a
description of the cases simulated. Numerical experiments, including a
Haines jump case, that show convergence and stability are given in
Section~\ref{sec:verification} and a comparison of the method
performances are made in
Section~\ref{sec:performance}. Section~\ref{sec:conclusion} summarizes
and concludes the paper.

\section{Pore network model}
\label{sec:model}

\begin{figure}[tbp]
  \centering
  \begin{subfigure}[b]{0.48\textwidth}
    \includegraphics[width=\textwidth]{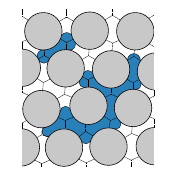}
    \caption{}
  \end{subfigure}
  ~
  \begin{subfigure}[b]{0.48\textwidth}
    \includegraphics[width=\textwidth]{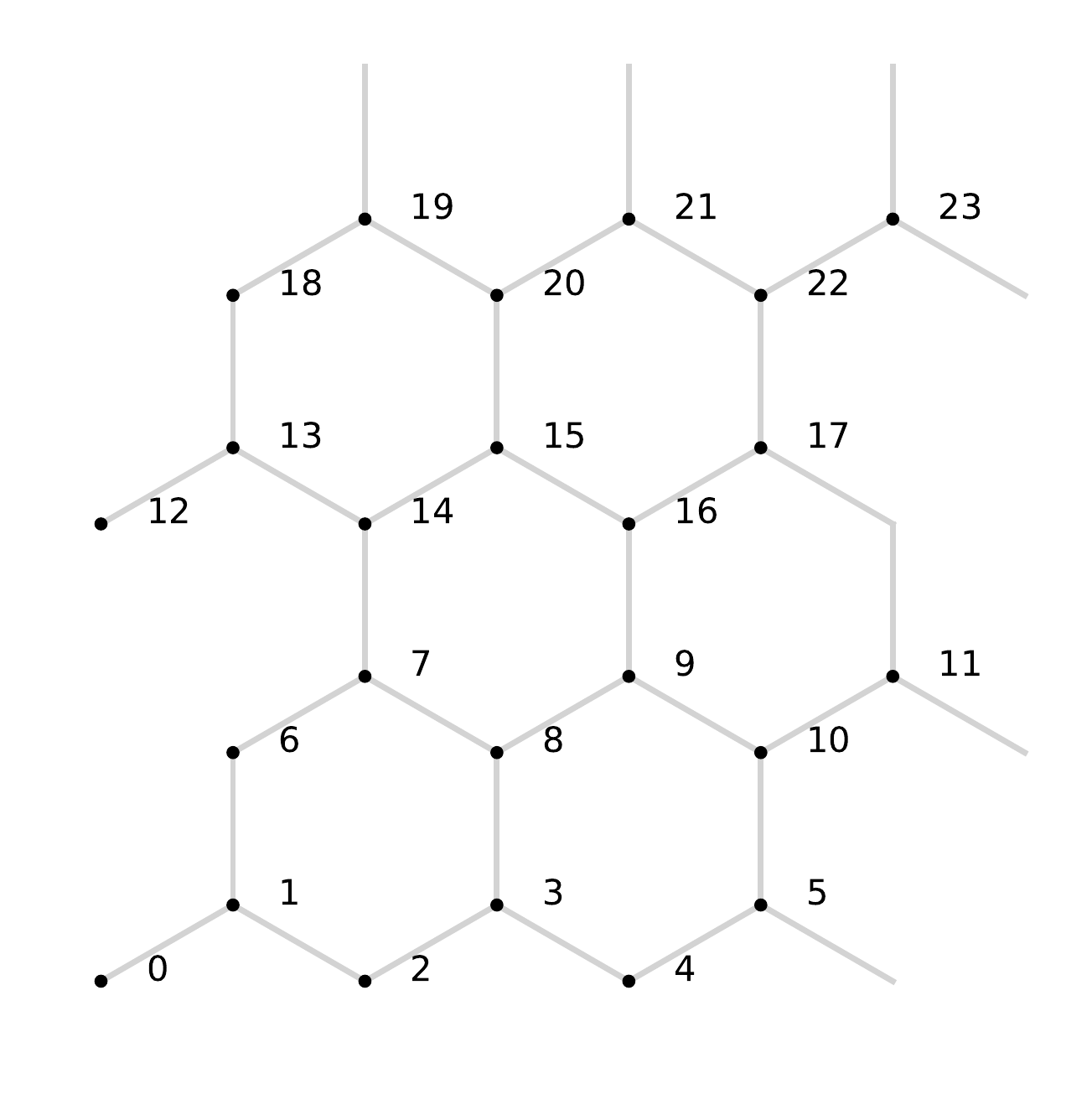}
    \caption{}
  \end{subfigure}
  \caption{Illustration of (a) a physical pore network with wetting
    (white) and non-wetting fluid (blue) and (b) its representation in
    the pore network model. The void space volumes separated by dashed
    lines in (a) are each represented as one link in (b). The node
    points in the model representation (b) is assumed to be located at
    the intersection points of the dashed lines in (a). Each fluid is
    assumed to fill the entire link cross section. The interface
    positions are therefore each represented in the model by a single
    number, giving its location along the link length.}
  \label{fig:network_model}
\end{figure}

We consider incompressible flow of two immiscible fluids in a porous
medium, where one fluid is more wetting towards the pore walls than
the other. We call the less wetting fluid non-wetting ($\n$) and the
more wetting fluid we call wetting ($\w$). The porous medium is
represented in the model by a network of $N$ nodes connected by $M$
links. Each node is given an index $i \in \left[0,N-1\right]$, and
each link is identified by the indices of the two nodes it
connects. An example pore network is shown in
Figure~\ref{fig:network_model}. The nodes are points that have no
volume and, consequently, all fluid is contained in the links. The
links therefore represent both the pore and the throat volumes of the
physical porous medium. In this respect, the pore network model
studied here differ from most other pore network models
\cite{Joekar-Niasar2012}. Each fluid is assumed to fill the entire
link cross section. The interface positions are therefore each
represented in the model by a single number, giving its location along
the link length.

The flow in each link is treated in a one-dimensional manner, where
the flow is averaged over the link cross section. As we consider flow in
relatively small cross sections only, we neglect any inertial effects
and the volume flow rate (\si{\meter\cubed\per\second}) from node $j$
to node $i$ through the link connecting then is given
by~\cite{Washburn1921}
\begin{linenomath}
  \begin{align}
    \label{eq:q_ij}
    q_{ij} &= -g_{ij} \left( \vec{z}_{ij} \right) \left\{ p_i - p_j -
    c_{ij} \left( \vec{z}_{ij} \right) \right\}.
  \end{align}
\end{linenomath}
Herein, $p_i$ (\si{\pascal}) is the pressure in node $i$, $g_{ij}$
(\si{\meter\cubed\per\second\per\pascal}) is the link mobility,
$c_{ij}$ (\si{\pascal}) is the link capillary pressure and
$\vec{z}_{ij}$ (\si{\meter}) is a vector containing the positions of
any fluid interfaces present in the link. Both the link mobility and
the capillary pressure depend on the fluid interface positions in the
link. If two nodes $i$ and $j$ are \textit{not} connected by a link,
then $g_{ij} = 0$. Due to mass conservation, the net flow rate into
every node $i$ is zero
\begin{linenomath}
  \begin{align}
    \label{eq:mass_conservation}
    \sum_j q_{ij} &= 0.
  \end{align}
\end{linenomath}
While the mobilities are symmetric with respect to permutation of the
indices, the capillary pressures are anti-symmetric,
\begin{linenomath}
  \begin{align}
    g_{ij} &= g_{ji}, \\
    c_{ij} &= -c_{ji}.
  \end{align}
\end{linenomath}
Introducing this into \eqref{eq:q_ij}, we obtain the immediately
intuitive result
\begin{linenomath}
  \begin{align}
  q_{ij} &= -q_{ji}.
  \end{align}
\end{linenomath}

The cross-sectional area of link $ij$ is denoted $a_{ij}$
(\si{\meter\squared}). Interface positions are advected with the flow
according to
\begin{linenomath}
  \begin{align}
    \label{eq:dzdt}
    \od{}{t} \vec{z}_{ij} = \frac{q_{ij}}{a_{ij}},
  \end{align}
\end{linenomath}
when they are sufficiently far away from the nodes. Near the nodes,
however, the interfaces are subject to additional modeling to account
for interface interactions in the pores. This is discussed further in
Section \ref{sec:bubble_rules}.

The form of the expressions for the mobilities and capillary pressures
depends on the shape of the links, and many different choices and
modeling approaches are possible. Here, we will use models similar to
those previously presented and used by e.g.\ \citet{Knudsen2002} and
\citet{Aker1998}. However, the treated time integration methods are
more general and can be applied to other models as well.

\subsection{Link mobility model}
\label{sec:g_model}

We apply a cylindrical link model when computing the mobilities, so
that
\begin{linenomath}
  \begin{align}
    g_{ij}(\vec{z}_{ij}) &= \frac{\pi r_{ij}^4}{8 L_{ij} \mu_{ij} \left(
      \vec{z}_{ij} \right)}.
  \end{align}
\end{linenomath}
Here, $r_{ij}$ (\si{\meter}) is the link radius and $L_{ij}$
(\si{\meter}) is the link length.  The viscosity $\mu_{ij}$
(\si{\pascal\second}) is the volume-weighted average of the fluid
viscosities and can be computed from the wetting and non-wetting fluid
viscosities $\mu_\w$ and $\mu_\n$ and the wetting fluid saturation
$s_{ij}$,
\begin{linenomath}
  \begin{align}
    \mu_{ij} \left( \vec{z}_{ij} \right) &= \mu_\w s_{ij}
    \left(\vec{z}_{ij}\right) + \mu_\n \left\{1 - s_{ij}
    \left(\vec{z}_{ij}\right) \right\}.
  \end{align}
\end{linenomath}

\subsection{Capillary pressure model}
\label{sec:c_model}

In each link $ij$, there may be zero, one or more interfaces
present. These are located at the positions specified in
$\vec{z}_{ij}$. As the interfaces may be curved, there may be a
discontinuity in pressure at these interface locations. The capillary
pressure $c_{ij}$ is the sum of interfacial pressure discontinuities
in the link $ij$. When computing the capillary pressures, we assume
that the links are wide near each end, and therefore that interfaces
located near a link end have negligible curvature and no pressure
discontinuity, while the links have narrow throats in the middle. The
link capillary pressures are thus modeled as
\begin{linenomath}
  \begin{align}
    \label{eq:c_ij}
    c_{ij} \left( \vec{z}_{ij} \right) &= \frac{2\sigma_{\w\n}}{r_{ij}}
    \sum_{z \in \vec{z}_{ij}} \left( \pm 1 \right) \left\{ 1 - \cos
    \left( 2 \pi \chi_{ij} \left(z\right) \right) \right\}.
  \end{align}
\end{linenomath}
The interfacial tension between the fluids is denoted $\sigma_{\w\n}$
(\si{\newton\per\meter}) and
\begin{linenomath}
  \begin{align}
    \chi_{ij} \left( z \right) =
    \begin{cases}
      0 & z < \alpha r_{ij}, \\ 
      \frac{z - \alpha r_{ij}}{L_{ij} -  2\alpha r_{ij}} &
      \alpha r_{ij} < z < L_{ij} - \alpha r_{ij}, \\
      1 & z > L_{ij} - \alpha r_{ij}.
    \end{cases}
  \end{align}
\end{linenomath}
The $\chi_{ij}$-function ensures zones of length $\alpha
r_{ij}$ at both ends of each link with zero capillary pressure across
any interface located there. Choosing $\alpha = 0$ is equivalent to
replacing $\chi_{ij}$ with $z/L_{ij}$ in \eqref{eq:c_ij}.

\subsection{Fluid interface interaction models}
\label{sec:bubble_rules}

The equations discussed so far in this section describe how the fluids
and the fluid interfaces move through the links. In addition, we rely
on models for how they behave close to the nodes. The purpose of these
are to emulate interface interactions in the pore spaces.

\begin{figure}[tbp]
  \centering
  \begin{subfigure}[b]{0.3\textwidth}
    \includegraphics[width=\textwidth]{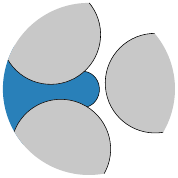}
    \caption{}
  \end{subfigure}
  ~
  \begin{subfigure}[b]{0.3\textwidth}
    \includegraphics[width=\textwidth]{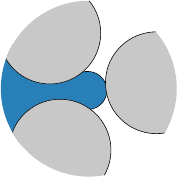}
    \caption{}
  \end{subfigure}
  \caption{Network node connected to three links. The node point,
    located near the middle of the pore space, is occupied by
    non-wetting fluid (blue). (a) The non-wetting fluid extends only a
    short distance into the links containing wetting fluid
    (white). The wetting fluid therefore remains connected and may
    flow freely through the pore space. (b) Non-wetting fluid
    protrudes far enough into all links to block the pore space for
    wetting fluid. The wetting fluid must now displace the non-wetting
    fluid in order to flow through.}
  \label{fig:flow_through}
\end{figure}

\begin{figure}[tbp]
  \centering
  \begin{subfigure}[b]{0.3\textwidth}
    \includegraphics[width=\textwidth]{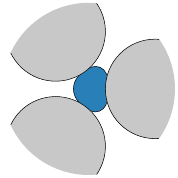}
    \caption{}
  \end{subfigure}
  ~
  \begin{subfigure}[b]{0.3\textwidth}
    \includegraphics[width=\textwidth]{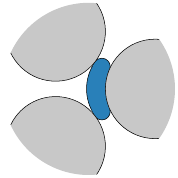}
    \caption{}
  \end{subfigure}
    ~
  \begin{subfigure}[b]{0.3\textwidth}
    \includegraphics[width=\textwidth]{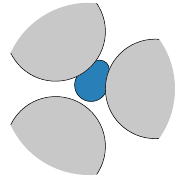}
    \caption{}
  \end{subfigure}
  \caption{(a) Small non-wetting bubble (blue) whose volume is small
    compared to the link volumes and is prevented from splitting by
    interfacial tension. This limits the minimum size of non-wetting
    bubbles, which will either (b) be stuck or (c) move through one of
    the links without splitting.}
  \label{fig:small_droplet}
\end{figure}

The following is assumed about the fluid behavior near the nodes and
is accounted for by the fluid interface interaction models.
\begin{itemize}
\item The mass of each fluid is conserved at every node. This means
  that at all times, all wetting and non-wetting fluid flowing into a
  node from one subset of its neighboring links must flow out into
  another disjoint subset of its neighboring links.
\item The network nodes in the model have no volume. However, due to
  the finite size of the physical pore void spaces, wetting fluid
  flowing into a pore space must be able to flow freely past any
  non-wetting fluid occupying the node point if the non-wetting fluid
  does not extend far enough into the pore void space cut the wetting
  fluid off. An example is illustrated in Figure
  \ref{fig:flow_through}. We consider a link $ij$ to be cut off from
  free outflow of wetting fluid if the non-wetting fluid continuously
  extends a length at least $\alpha r_{ij}$ into the link. Non-wetting
  fluid may freely flow past wetting fluid, or not, the same manner.
\item In each link $ij$, interfacial tension will prevent droplets
  with length smaller than $\alpha r_{ij}$ from forming by separation
  from larger droplets. An example is illustrated in Figure
  \ref{fig:small_droplet}.
\end{itemize}

\subsection{Boundary conditions}

We consider only networks where the nodes and links can be laid out in
the two-dimensional $x$-$y$ plane. These networks will be periodic in
both the $x$- and $y$-direction. However, the model is also applicable
to networks that extend in three dimensions \cite{Sinha2017}, and the
presented numerical methods are also compatible both with networks in
three dimensions and with other, non-periodic boundary conditions
\cite{Erpelding2013}.

We will here apply two types of boundary conditions to the
flow. With the first type, a specified pressure difference $\Delta P$
(\si{\pascal}) will be applied across the network in the
$y$-direction. This pressure difference will be equal to the sum of
all link pressure differences in any path spanning the network once in
the $y$-direction, ending up in the same node as it started. With the
other type of boundary condition, we specify a total flow rate $Q$
(\si{\meter\cubed\per\second}) across the network. This flow rate will
be equal to the sum of link flow rates flowing through any plane drawn
through the network normal to the $y$-axis.

\section{Temporal discretization}
\label{sec:methods}

In the following three sections, we describe the different time
integration methods considered. These methods are applied to
\eqref{eq:dzdt}, where evaluation of the right hand side involves
simultaneously solving the mass conservation equations
\eqref{eq:mass_conservation} and the constitutive equations
\eqref{eq:q_ij} to obtain all unknown link flow rates and node
pressures.

The discretized times (\si{\second}) are denoted with a superscript
where $n$ is the time step number,
\begin{linenomath}
  \begin{align}
    t^{(n)} &= t^{(0)} + \sum_{i=0}^{n-1} \Delta t^{(i)}.
  \end{align}
\end{linenomath}
The time step $\Delta t^{(i)}$ is the difference between $t^{(i+1)}$
and $t^{(i)}$ and the time $t^{(0)}$ is the initial time in a
simulation. Similarly, quantities evaluated at the discrete times are
denoted with time step superscripts, e.g.
\begin{linenomath}
  \begin{align}
    q_{ij}^{(n)} &= q_{ij} \left( t^{(n)} \right).
  \end{align}
\end{linenomath}
Mobilities and capillary pressures with superscripts are evaluated
using the interface positions at the indicated time step,
\begin{linenomath}
  \begin{align}
    g_{ij}^{(n)} &= g_{ij}\left( \vec{z}_{ij}^{(n)}\right), \\
    c_{ij}^{(n)} &= c_{ij}\left( \vec{z}_{ij}^{(n)}\right).
  \end{align}
\end{linenomath}

\section{Forward Euler method}
\label{sec:fem}

The forward Euler method is the simplest of the time integration
methods considered here and is the one used most frequently in
previous works, see e.g.~\citep{Knudsen2002} and \citep{Sinha2012}. We
include its description here for completeness and to provide context
for the proposed new capillary time step criterion that is introduced
to stabilize the method at low capillary numbers.

The ordinary differential equation (ODE) \eqref{eq:dzdt} is
discretized in a straightforward manner for each link $ij$ using
forward Euler,
\begin{linenomath}
  \begin{align}
    \label{eq:dzdt_fem}
    \vec{z}_{ij}^{(n+1)} &= \vec{z}_{ij}^{(n)} + \Delta t^{(n)}
    \frac{q_{ij}^{(n)}}{a_{ij}}.
  \end{align}
\end{linenomath}  
The flow rates are calculated by inserting \eqref{eq:q_ij}, evaluated
with the current known interface positions,
\begin{linenomath}
  \begin{align}
    \label{eq:q_ij_explicit}
    q_{ij}^{(n)} &= -g_{ij}^{(n)} \left\{ p_i^{(n)} - p_j^{(n)} - c_{ij}^{(n)} \right\},
  \end{align}
\end{linenomath}
into the mass conservation equations
\eqref{eq:mass_conservation}. This results in the a system of linear
equations consisting of one equation,
\begin{linenomath}
  \begin{align}
    \sum_j g_{ij}^{(n)} p_j^{(n)} - p_i^{(n)} \sum_j g_{ij}^{(n)}
    = - \sum_j g_{ij}^{(n)} c_{ij}^{(n)},
  \end{align}
\end{linenomath}
for each node $i$ with unknown pressure. This linear system can be
cast into matrix form,
\begin{linenomath}
  \begin{align}
    \label{eq:fem_linear_system}
    \mat{A} \cdot \vec{x} = \vec{b},
  \end{align}
\end{linenomath}
where the vector \vec{x} contains the unknown node pressures, e.g.
\begin{linenomath}
  \begin{align}
    \vec{x} = 
    \begin{bmatrix}
      p_0^{(n)} \\
      p_1^{(n)} \\
      \vdots \\
      p_{N-1}^{(n)}
    \end{bmatrix}.
  \end{align}
\end{linenomath}
The matrix elements are
\begin{linenomath}
  \begin{align}
    A_{ij} = \left\{ 1 - \delta_{ij} \right\} g_{ij}^{(n)} - \delta_{ij}
    \sum_k g_{ik}^{(n)},
  \end{align}
\end{linenomath}
and the elements of the constant vector are
\begin{linenomath}
  \begin{align}
    b_{i} = -\sum_k g_{ik}^{(n)} c_{ik}^{(n)}.
  \end{align}
\end{linenomath}
The node pressures are obtained by solving this linear equation
system. The flow rates are subsequently evaluated using
\eqref{eq:q_ij_explicit} and the interface positions are then updated
using \eqref{eq:dzdt_fem} and the interface interaction models.

\subsection{Time step restrictions}
\label{sec:fem_dt}

In previous works \cite{Aker1998,Knudsen2002}, the time step length
was chosen from a purely advective criterion,
\begin{linenomath}
  \begin{align}
    \label{eq:dt_a}
    \Delta t^{(n)}_\advective = C_\advective \min_{ij} \left(
    \frac{a_{ij} L_{ij}}{q_{ij}^{(n)}} \right).
  \end{align}
\end{linenomath}
The parameter $C_\advective$ corresponds to the maximum fraction of a
link length any fluid interface is allowed to move in a single forward
Euler time step. The value of $C_\advective$ must be chosen based on
the level of accuracy desired from the simulation.

However, selecting the time step based on the advective criterion
only, often results in numerical instabilities at low capillary
numbers, where viscous forces are small relative to the capillary
forces. This is demonstrated in Section \ref{sec:stability_tests}. The
origins of the numerical instabilities can be identified by performing
analysis on a linearized version of the governing equations. This is
done in \ref{sec:capillary_time_step}. This analysis also leads to a
new time step criterion, whereby the time step length is restricted by
the sensitivity of the capillary forces to perturbations in the
current interface positions,
\begin{linenomath}
  \begin{align}
    \label{eq:dt_c}
    \Delta t^{(n)}_\capillary &= C_\capillary 
    \min_{ij} \left( \frac{2 a_{ij} }{ g_{ij}^{(n)} \left| \sum_{z \in \vec{z}_{ij}^{(n)}}
      \pd{c_{ij}}{z} \right|} \right).
  \end{align}
\end{linenomath}
For the particular choice of capillary pressure model given by
\eqref{eq:c_ij}, we obtain
\begin{linenomath}
  \begin{align}
    \Delta t^{(n)}_\capillary &= C_\capillary \min_{ij}
    \left( \frac{a_{ij} r_{ij} L_{ij}}{ 2 \pi g_{ij}^{(n)} \sigma_{\w\n} \left|
      \sum_{z \in \vec{z}_{ij}^{(n)}} \left( \pm 1 \right) \sin \left( 2
      \pi \chi_{ij} \left(z \right) \right) \od{\chi_{ij}}{z} \right| } \right).
  \end{align}
\end{linenomath}
According to the linear analysis, numerical instabilities are avoided
if the parameter $C_\capillary$ is chosen such that $0 < C_\capillary
< 1$. However, we must regard \eqref{eq:dt_c} as an approximation when
we apply it to the full non-linear model in simulations and,
consequently, we may have to chose $C_\capillary$ conservatively to
ensure stability for all cases.

At each step in the simulation, the time step used is then taken as
\begin{linenomath}
  \begin{align}
    \Delta t^{(n)} = \min \left( \Delta t^{(n)}_\capillary, \Delta
    t^{(n)}_\advective \right),
  \end{align}
\end{linenomath}
to comply with both the advective and the capillary time step
criteria. The capillary time step restriction \eqref{eq:dt_c} is
independent of flow rate. It therefore becomes quite severe, demanding
relatively fine time steps, when flow rates are low.

\subsection{Boundary conditions}
\label{sec:fem_bcs}

The periodic boundary conditions, specifying a total pressure
difference $\Delta P$ across the network, can be incorporated directly
into the linear equation system \eqref{eq:fem_linear_system}. For each
node $i$, a term $g_{ij}^{(n)} \Delta P$ is added to or subtracted
from $b_i$ for any link $ij$ that crosses the periodic boundary.

With the specified $\Delta P$ condition implemented, we can use it to
obtain the node pressures and link flow rates corresponding to a
specified total flow rate $Q$. Due to the linear nature of the model,
the total flow rate is linear in $\Delta P$ \cite{Aker1998}, so that
\begin{linenomath}
  \begin{align}
    \label{eq:linear_bc}
    Q = C_1 \Delta P + C_2,
  \end{align}
\end{linenomath}
for some unknown coefficients $C_1$ and $C_2$, that are particular to
the current fluid configuration.

We choose two different, but otherwise arbitrary, pressure drop values
$\Delta P_1$ and $\Delta P_2$ and, using the above procedure, we solve
the network model once for each pressure difference and calculate the
corresponding total flow rates $Q_1$ and $Q_2$. The coefficients $C_1$
and $C_2$ are then determined by,
\begin{linenomath}
  \begin{align}
    C_1 &= \frac{Q_2 - Q_1}{\Delta P_2 - \Delta P_1}, \\
    C_2 &= \frac{Q_2 \Delta P_1 - Q_1 \Delta P_2}{\Delta P_1 - \Delta P_2}.
  \end{align}
\end{linenomath}
The pressure difference $\Delta P$ required to obtain the specified
flow rate $Q$ is determined by solving \eqref{eq:linear_bc} for
$\Delta P$. Subsequently, the network model is solved a third time
with pressure drop $\Delta P$ to obtain the desired node pressures and
link flow rates.

\section{Midpoint method}
\label{sec:mpm}

The forward Euler method is first-order accurate in time. To obtain
smaller numerical errors, methods of higher order are desirable. We
therefore include in our discussion the second-order midpoint
method. This method is identical to that used by \citet{Aker1998},
except with respect to choice of time step length.

The ODE \eqref{eq:dzdt} is discretized as
\begin{linenomath}
  \begin{align}
    \label{eq:dzdt_mpm}
    \vec{z}^{(n+1)}_{ij} &= \vec{z}^{(n)}_{ij} + \Delta t^{(n)}
    \frac{q_{ij}^{(n+1/2)}}{a_{ij}},
  \end{align}
\end{linenomath}
where $q_{ij}^{(n+1/2)}$ is the flow rate at the midpoint in time
between point $n$ and $n+1$. This flow rate is calculated in the same
manner as described in Section~\ref{sec:fem}. The interface positions
at $n + 1/2$ are obtained by taking a forward Euler step with half the
length of the whole time step,
\begin{linenomath}
  \begin{align}
    \vec{z}^{(n+1/2)}_{ij} &= \vec{z}^{(n)}_{ij} + \frac{1}{2} \Delta t^{(n)}
    \frac{q_{ij}^{(n)}}{a_{ij}}.
  \end{align}
\end{linenomath}

\subsection{Time step restrictions}
Since the forward Euler stability region is contained within the
stability region for the midpoint method, we use the same time step
restrictions for the midpoint method as for forward Euler, see
Section~\ref{sec:fem_dt}.

\subsection{Boundary conditions}

Both the specified $\Delta P$ and the specified $Q$ boundary
conditions are incorporated into the midpoint method by applying the
procedures described in Section \ref{sec:fem_bcs} for each evaluation
of the right hand side of \eqref{eq:dzdt}.

\section{Semi-implicit method}
\label{sec:sim}

To avoid both the numerical instabilities and the time step
restriction~\eqref{eq:dt_c}, which becomes quite severe at low flow
rates, we here develop a new semi-implicit time stepping
method. Simulation results indicate that this method is stable with
time steps determined by the advective criterion \eqref{eq:dt_a} only,
and much longer time steps are therefore possible than with the
forward Euler and midpoint methods at low capillary numbers.

The ODE~\eqref{eq:dzdt} is now discretized according to
\begin{linenomath}
  \begin{align}
    \label{eq:dzdt_sim}
    \vec{z}_{ij}^{(n+1)} &= \vec{z}_{ij}^{(n)} + \Delta t^{(n)}
    \frac{q_{ij}^{(n+1)}}{a_{ij}}.
  \end{align}
\end{linenomath}  
The semi-implicit nature of this discretization comes from the flow
rate used,
\begin{linenomath}
  \begin{align}
    \label{eq:q_ij_sim}
    q_{ij}^{(n+1)} &= -g_{ij}^{(n)} \left\{ p_i^{(n+1)} - p_j^{(n+1)} -
    c_{ij}^{(n+1)} \right\}.
  \end{align}
\end{linenomath}
Herein, the link mobility is evaluated at time step $n$, while the
node pressures and the capillary pressure are evaluated time step
$n+1$.

The link mobilities could of course also have been evaluated at time
step $n+1$, thus creating a fully implicit backward Euler scheme. As
is shown in \ref{sec:capillary_time_step}, we may expect backward
Euler to be stable with any positive $\Delta t^{(n)}$. The backward
Euler scheme may therefore seem like a natural choice for performing
stable simulations with long time steps. However, to evaluate the
mobilities at time step $n+1$ complicates the integration procedure
and was found to be unnecessary in practice. A semi-implicit
alternative is therefore preferred.

To obtain the node pressures, we solve the mass conservation equations,
\begin{linenomath}
  \begin{align}
    F_i &= \sum_k q_{ik}^{(n+1)} = 0.
  \end{align}
\end{linenomath}
Again, we have one equation for each node $i$ with unknown pressure.
However, because the capillary pressures now depend on the flow rates,
\begin{linenomath}
  \begin{align}
    c_{ij}^{(n+1)} &= c_{ij} \left( \vec{z}_{ij}^{(n)} + \Delta t^{(n)}
    \frac{q_{ij}^{(n+1)}}{a_{ij}} \right),
  \end{align}
\end{linenomath}
the mass conservation equations are now a system of non-linear
equations, rather than a system of linear equations. This system can
be cast in the form
\begin{linenomath}
  \begin{align}
    \label{eq:non-linear_system}
    \vec{F} \left( \vec{x} \right) = \vec{0},
  \end{align}
\end{linenomath}
where $\vec{x}$ contains the unknown pressures, e.g.
\begin{linenomath}
  \begin{align}
    \vec{x} &= 
    \begin{bmatrix}
      p_0^{(n+1)} \\
      p_1^{(n+1)} \\
      \vdots \\
      p_{N-1}^{(n+1)}
    \end{bmatrix}.
  \end{align}
\end{linenomath}

In order to solve \eqref{eq:non-linear_system} using the numerical
method described in Section \ref{sec:implementation}, it is necessary
to have the Jacobian matrix of $\vec{F}$. Details on how the Jacobian
matrix is calculated are given in \ref{sec:jacobian}.

The calculation of link flow rates from node pressures, and thus every
evaluation of $\vec{F}$ and its Jacobian, involves solving one
non-linear equation for each link flow rate,
\begin{linenomath}
  \begin{align}
    \label{eq:sim_G_ij}
    G_{ij} \left( q_{ij}^{(n+1)} \right) = q_{ij}^{(n+1)} + g_{ij}^{(n)}
    \left\{ p_i^{(n+1)} - p_j^{(n+1)} - c_{ij}^{(n+1)} \right\} = 0.
  \end{align}
\end{linenomath}
The derivative of $G_{ij}$ with respect to $q_{ij}^{(n+1)}$ is
\begin{linenomath}
  \begin{align}
    \od{G_{ij}}{q_{ij}^{(n+1)}} = 1 - g_{ij}^{(n)} \od{c_{ij}^{(n+1)}}{q_{ij}^{(n+1)}}.
  \end{align}
\end{linenomath}

The procedure for updating the interface positions with the
semi-implicit method may be summarized as follows. The non-linear
equation system \eqref{eq:non-linear_system} is solved to obtain the
unknown node pressures. In every iteration of the solution procedure,
the flow rates are evaluated by solving \eqref{eq:sim_G_ij} for each
link. When a solution to \eqref{eq:non-linear_system} is obtained, the
interface positions are updated using \eqref{eq:dzdt_sim} and the
interface interaction models.

\subsection{Time step restrictions}
\label{sec:sim_dt}

We aim to select the time steps such that
\begin{linenomath}
  \begin{align}
    \Delta t^{(n)} = \Delta t^{(n+1)}_\advective.
  \end{align}
\end{linenomath}
However, to solve the non-linear system \eqref{eq:non-linear_system}
is challenging in practice and requires initial guess values for the
link flow rates and node pressures that lie sufficiently close to the
solution. For this purpose, we here use values from the previous time
step. This turns out to be a sufficiently good choice for most time
steps, but our numerical solution procedure does not always
succeed. As the link flow rates and node pressures at two consecutive
points in time become increasingly similar as the time interval
between them is reduced, we may expect the guess values to lie closer
to the solution if we reduce the time step. Thus, if our solution
procedure is unable to succeed, our remedy is to shorten $\Delta
t^{(n)}$. This will sometimes lead to time steps shorter than $\Delta
t^{(n+1)}_\advective$. If, for a given time step, $\Delta t^{(n)}$
must be reduced to less than twice the time step length allowed by the
explicit methods, we revert to forward Euler for that particular
step. As we demonstrate in Section \ref{sec:performance}, however,
this does not prevent the semi-implicit method from being much more
efficient than the explicit methods at low capillary numbers.

\subsection{Boundary conditions}

As with the explicit methods, the specified $\Delta P$ boundary
condition can be incorporated directly into the mass balance equation
system, in this case \eqref{eq:non-linear_system}. This is done by
adding to or subtracting from the right hand sides of
\eqref{eq:q_ij_sim} and \eqref{eq:sim_G_ij} a term $g_{ij}^{(n)}
\Delta P$ for each link $ij$ crossing the periodic boundary.

The specified flow rate boundary condition is incorporated by
including $\Delta P$ as an additional unknown and adding an additional
equation
\begin{linenomath}
  \begin{align}
    \label{eq:sim_Q_BC}
    F_m &= \left\{ \sum_{ij \in \Omega} q_{ij}^{(n+1)} \right\} - Q = 0,
  \end{align}
\end{linenomath}
to the non-linear equation system
\eqref{eq:non-linear_system}. Herein, $\Omega$ is the set of links
crossing the periodic boundary, with $i$ being the node on the
downstream side and $j$ being the node on the upstream side. Thus,
\eqref{eq:sim_Q_BC} is satisfied when the total flow rate through the
network is equal to $Q$.

\section{Implementation}
\label{sec:implementation}

The non-linear equation system \eqref{eq:non-linear_system} is solved
using a Newton-type solution method that guarantees convergence to a
local minimum of $\vec{F} \cdot \vec{F}$, see
\citet{Press2007}~pp.\ 477. However, a local minimum of $\vec{F} \cdot
\vec{F}$ is not necessarily a solution to
\eqref{eq:non-linear_system}, and good initial guess values for the
node pressures and link flow rates are therefore crucial. For this
purpose, we use the values from the previous time step and reduce the
length of the current time step if the solution method fails, as
discussed in Section \ref{sec:sim_dt}.

Solving \eqref{eq:sim_G_ij} is done using a standard Newton solver
\cite{Suli2006}. For robustness, a bisection solver \cite{Suli2006} is
used if the Newton solver fails.

The Newton-type solver for non-linear systems and the explicit time
integration methods require methods for solving linear systems of
equations. We use the conjugate gradient method in combination with
the LU preconditioner implemented in the PETSc library, see
\citet{Balay2016}. An introduction to solving systems of
Kirchhoff-type equations numerically can be found in
\cite{Batrouni1988}.

\section{Case descriptions}
\label{sec:cases}

In this section, we describe the two simulated cases. One is a test
case where a single bubble is contained in a network consisting of
links connected in series, while the other is designed to capture a
single Haines jump in a small network where fluids flow at a specified
rate.

\subsection{Links-in-series test case}

The verification will include comparison of results from the various
numerical methods applied to a test case. The test case is chosen such
that it can be set up as a single ODE with a closed expression for the
right-hand side. An accurate reference solution can thus be easily
obtained using a high-order Runge--Kutta method. As our test case, we
consider a network consisting of $M=3$ identical links connected in
series. The network contains a single bubble of length $\ell$
(\si{\meter}) with center position $z$ (\si{\meter}). In the capillary
pressure model, we choose $\alpha = 0$. The ODE \eqref{eq:dzdt} can
then be restated as an equivalent equation for the bubble position,
\begin{linenomath}
  \begin{align}
    \od{z}{t} &= \frac{Q}{a},
  \end{align}
\end{linenomath}
where $Q$ is the flow through the network and $a$ is the link
cross-sectional area. The model equations can be reduced to the
following expression for flow rate.
\begin{linenomath}
  \begin{align}
    Q = -\frac{g}{M} \left\{ \Delta P + \frac{4\sigma_{\w\n}}{r}
    \sin\left( \frac{\pi \ell}{L}\right) \sin\left( \frac{2 \pi
      z}{L}\right) \right\}
  \end{align}
\end{linenomath}
Here, $g$ is the mobility of a single link, $L=\SI{1.0e-3}{\meter}$ is
the length of a single link and $r=\SI{1.0e-4}{\meter}$ is the link
radius. The bubble has length $\ell=\SI{4.8e-4}{\meter}$ and is
initially located at $z=\SI{2.4e-4}{\meter}$. The fluid parameters
used in all simulations are given in Table
\ref{tab:reference_params}. The pressure difference $\Delta P$ will be
stated for each simulation.

\begin{table}[tbp]
  \caption{Fluid properties corresponding to water (\w) and decane (\n) at
    atmospheric pressure and \SI{298}{\kelvin}}
  \label{tab:reference_params}
  \centering
  \begin{tabular}{c c c c}
    \toprule
    Parameter & Value & Unit & Reference \\
    \midrule
    $\mu_\w$ & \SI{8.9e-4} & \si{\pascal\second}
    & \cite{NISTChemistryWebBook} \\
    $\mu_\n$ & \SI{8.5e-4}{} & \si{\pascal\second}
    & \cite{NISTChemistryWebBook}\\
    $\sigma_{\w\n}$ & \SI{5.2e-2}{} & \si{\newton\per\meter}
    & \cite{Zeppieri2001}\\
    \bottomrule
  \end{tabular}
\end{table}

\subsection{Haines jump case}

The Haines jump was first reported almost 90 years ago
\cite{Haines1930}. It refers to the sudden drops in driving pressure
observed in drainage experiments when non-wetting fluid breaks through
a throat and invades new pores. This process was studied
experimentally and numerically by \citet{Maloy1992} and, more
recently, it was imaged directly and analyzed in detail by
\citet{Armstrong2013} for flow in a micromodel and by \citet{Berg2013}
for flow in a sample of Berea sandstone. The Haines jump case
simulated here captures one such break-through and subsequent pressure
drop.

Among the findings in the study by \citet{Maloy1992} was that pore
drainage is a non-local event, meaning that as one pore is drained,
imbibition occurs in nearby neck regions. This was also observed by
\citet{Armstrong2013}, and was explained as follows. When the imposed
flow rates are low, the non-wetting fluid that fills the newly invaded
pores needs to be supplied from nearby locations rather than the
external feed. \citet{Armstrong2013} also found, for their range of
investigated parameters, that pore drainage occurred on the same
time-scale, regardless of the externally imposed flow rate.

\begin{figure}[tbp]
  \centering
  \includegraphics[width=0.48\textwidth]{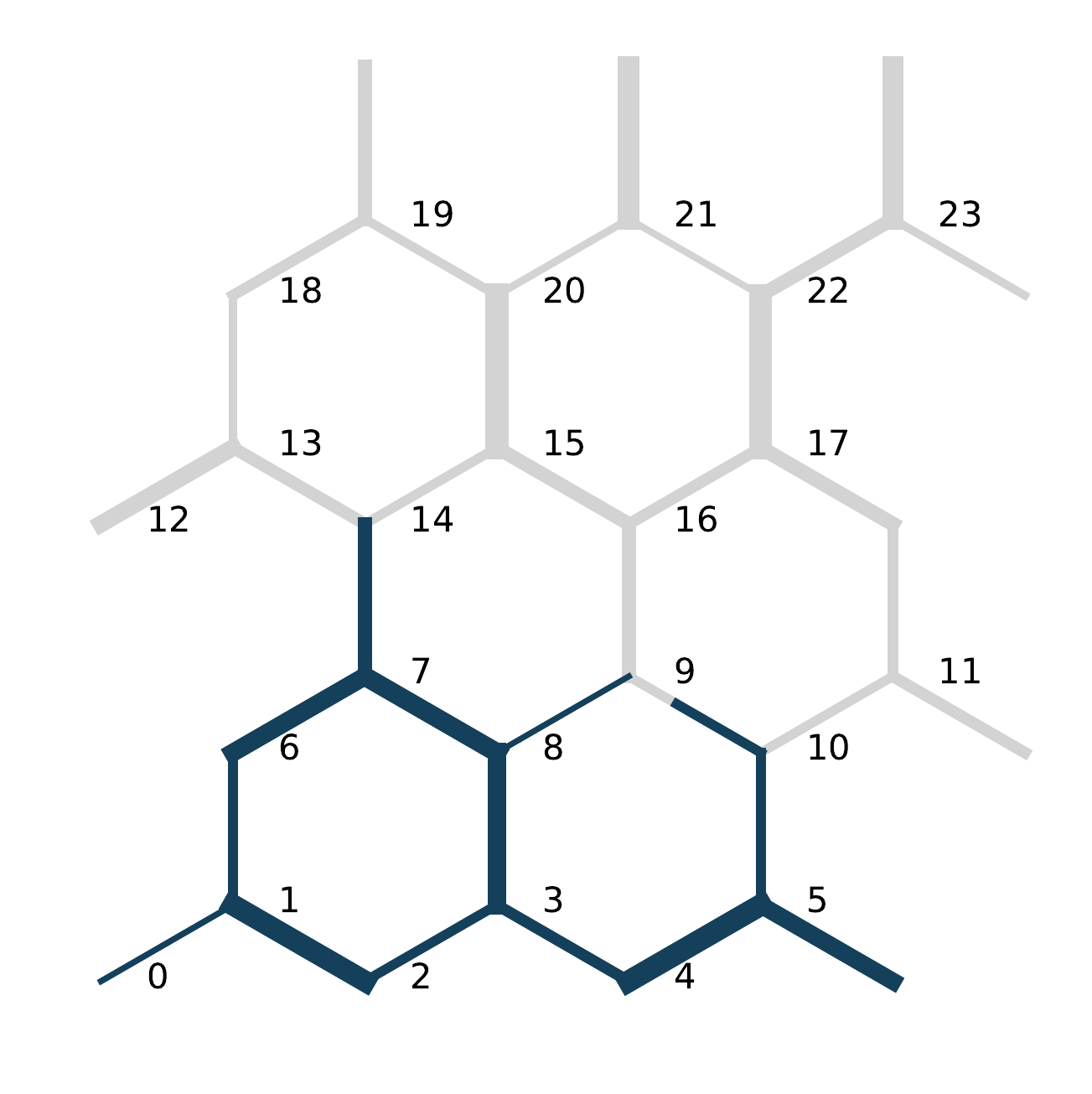}
  \caption{Initial fluid configuration in the Haines jump case. The
    non-wetting fluid is blue while the wetting fluid is gray. The
    link radii are not drawn to scale with the link lengths. Node
    indices are indicated in black.}
  \label{fig:network_initial}
\end{figure}

We consider a hexagonal network consisting $N=24$ nodes and $M=36$
links. All links have length \SI{1.0e-3}{\meter}, while the link radii
are drawn randomly from a uniform distribution between 0.1 and 0.4
link lengths. In the capillary pressure model, we choose $\alpha =
1$. The fluid parameters $\mu_\w$, $\mu_\n$ and $\sigma_{\w\n}$ are
the same as in the links-in-series test case, see Table
\ref{tab:reference_params}. With these fluid parameters and network
length scales, the case mimics the flow of water (\w) and decane (\n)
in a Hele-Shaw cell filled with glass beads similar to those used in
e.g.\ \cite{Maloy1985,Maloy1992,Tallakstad2009}. The linear dimensions
are $\sim10$ times bigger in this network compared to the micromodel
of \citet{Armstrong2013}. Initially, the fluids are distributed in the
network as shown in Figure \ref{fig:network_initial}, with the
non-wetting fluid in a single connected ganglion.

Simulations are run at different specified flow rates $Q$ until a net
fluid volume equivalent to \SI{5}{\percent} of the total pore volume
has flowed through the network. The flow dynamics will, of course,
depend upon the specified flow rate. At low flow rates, however, the
flow will exhibit some relatively fast fluid redistribution events and
one relatively slow pressure build-up and subsequent Haines jump
event. The Haines jump will occur as the non-wetting fluid breaks
through the link connecting nodes 9 and 16 and invades node 16, see
Figure~\ref{fig:network_initial}.

It was mentioned by \citet{Armstrong2013} that the large local flow
velocities that they observed as a pore was filled with non-wetting
fluid during a Haines jump has implications for how such processes
must be numerically simulated. Specifically, the time resolution of
the simulation needs to be fine enough during these events to capture
them. This poses a challenge when externally applied flow rates are
low and there is thus a large difference in the large time scale that
governs the overall flow of the system and the small time scale than
governs the local flow during Haines jumps.

\section{Verification}
\label{sec:verification}

In this section, we verify that the time integration methods presented
correctly solve the pore network model equations and that the time
step criteria presented give stable solutions.

\subsection{Convergence tests}
\label{sec:convergence_tests}

All time integration methods presented should, of course, give the
same solution for vanishingly small time steps. Furthermore, the
difference between the solution obtained with a given finite time step
and the fully converged solution should decrease as the time steps are
refined, and should do so at a rate that is consistent with the order
of the method. In this section, we verify that all three time
integration methods give solutions that converge to the reference
solution for the links-in-series test case and thus that the methods
correctly solve the model equations for this case.

\begin{figure}[p]
    \centering
    \begin{subfigure}[b]{0.8\textwidth}
      \includegraphics[width=\textwidth]{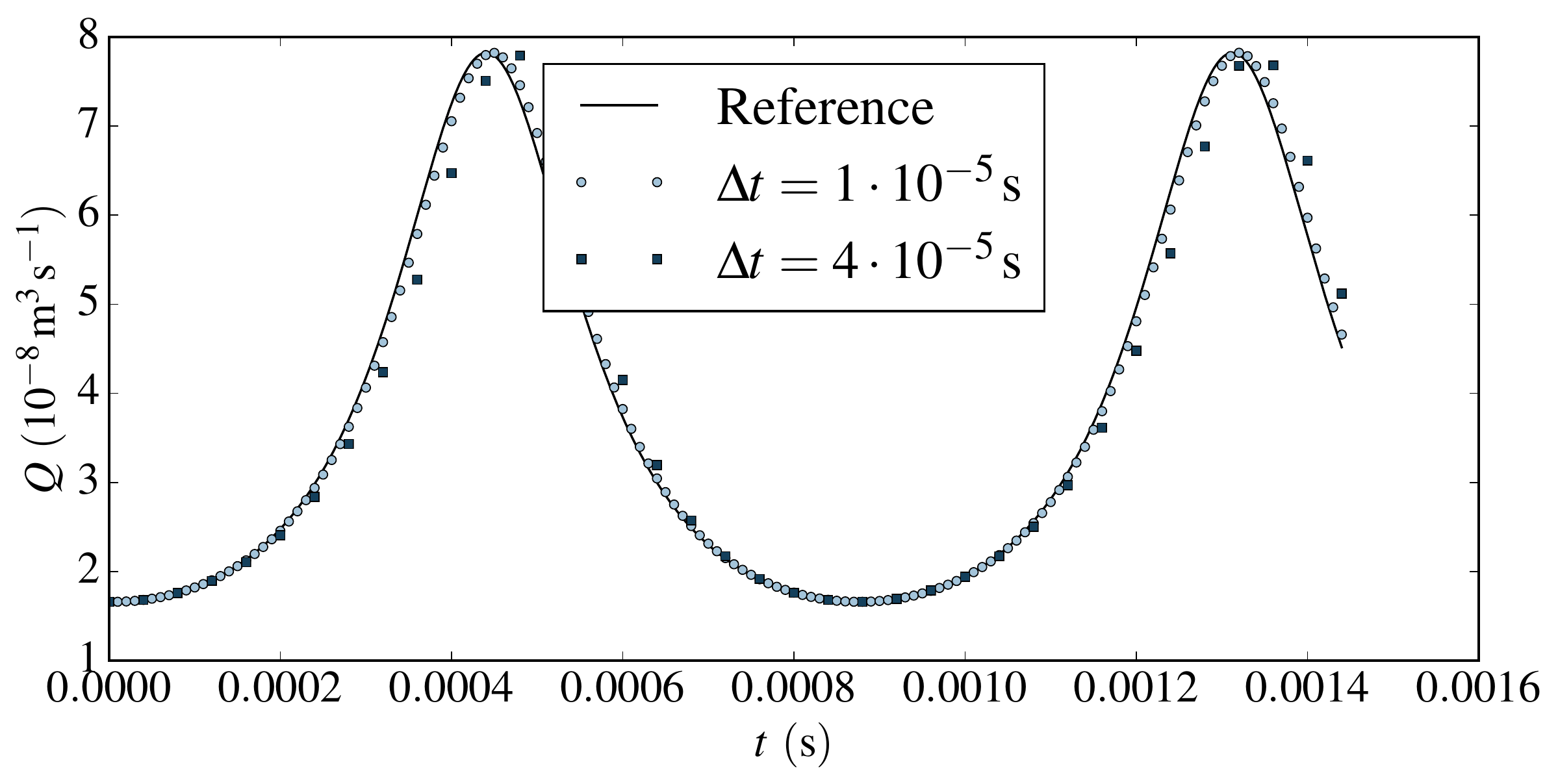}
      \caption{}
      \label{fig:Q_test_case_fem}
    \end{subfigure}
    \begin{subfigure}[b]{0.8\textwidth}
      \includegraphics[width=\textwidth]{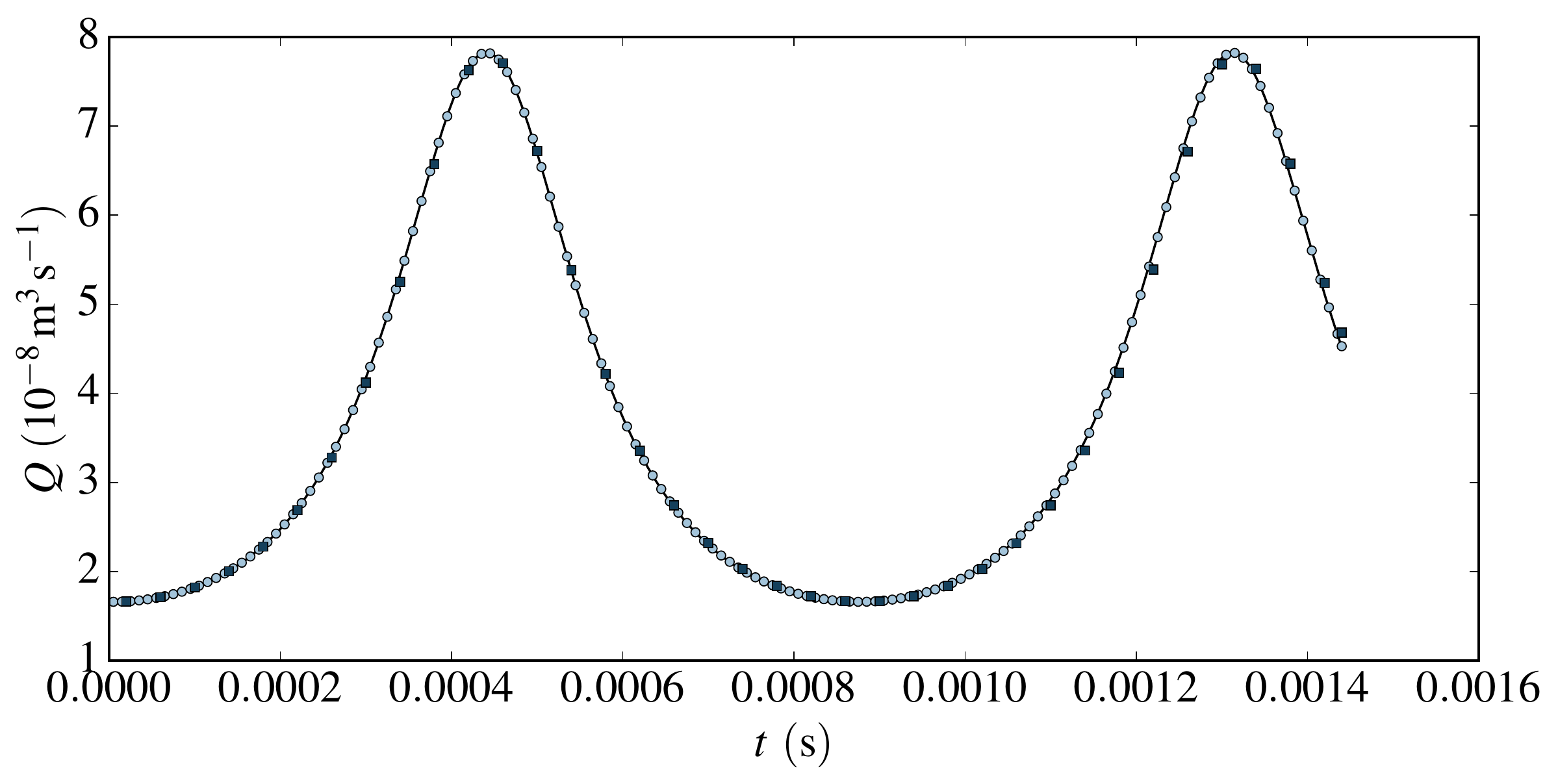}
      \caption{}
    \end{subfigure}
    \begin{subfigure}[b]{0.8\textwidth}
        \includegraphics[width=\textwidth]{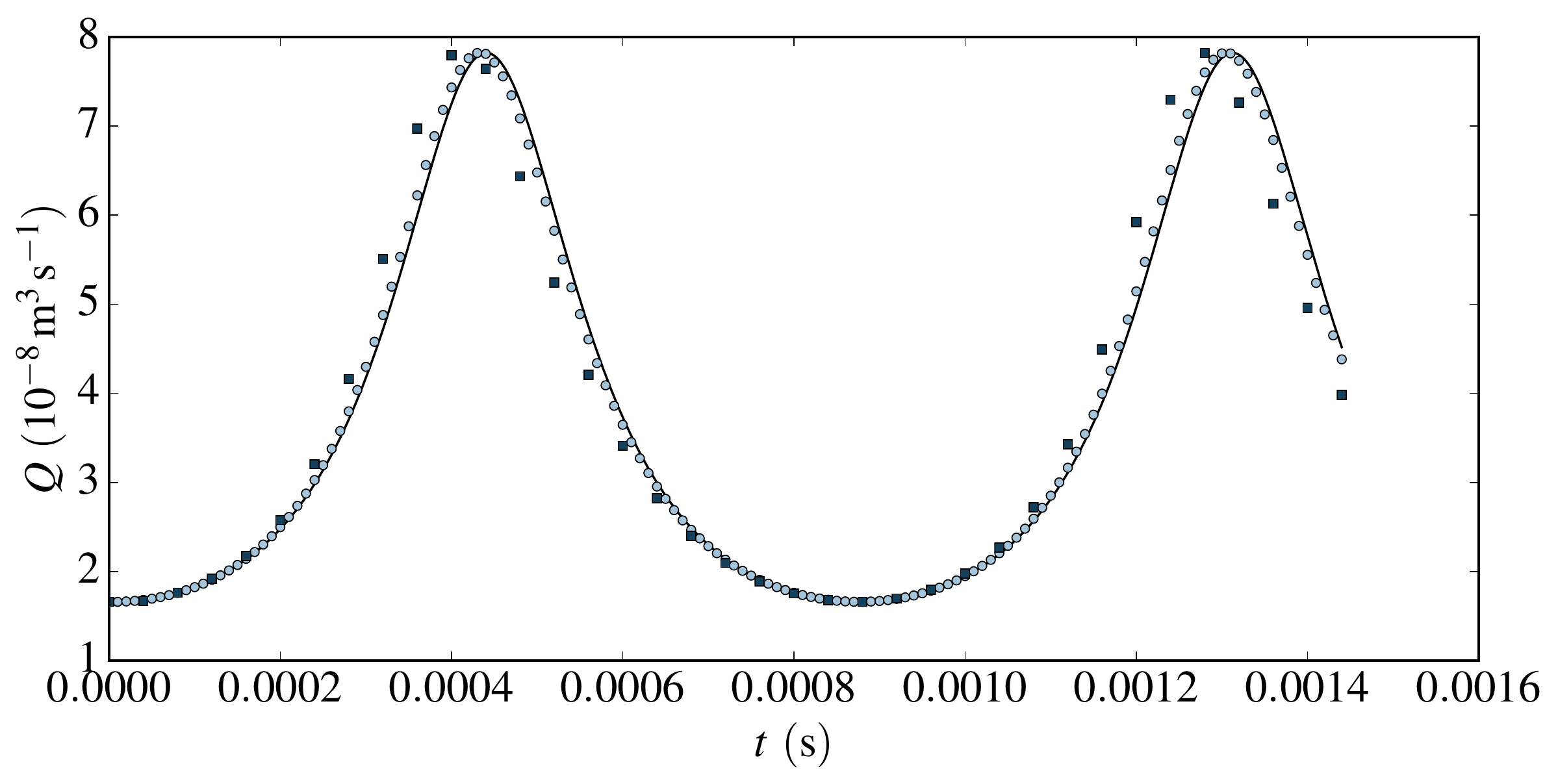}
        \caption{}
    \end{subfigure}
    \caption{Flow rates $Q$ plotted against time for two different
      time steps $\Delta t$ for the links-in-series test case with
      $\Delta P = -\SI{3200}{\pascal}$. Results from the forward Euler
      method are given in (a), results from the midpoint method in (b)
      and results from the semi-implicit method in (c). The solid line
      represents the reference solution.}
    \label{fig:z_Q_test_case}
\end{figure}

\begin{table}[p]
  \caption{Relative errors in bubble position $z$ and flow rate $Q$ at
    $t = \SI{0.00144}{\second}$ and estimated convergence orders for
    the links-in-series test case computed with the forward Euler
    method} \centering
  \begin{tabular}{c c c c c}

    \toprule
    $\Delta t$ $(\si{\second})$ & $z$-error & $z$-order & $Q$-error & $Q$-order \\
    \midrule
    \SI{ 4e-05 }{} & \SI{ 1.55e-02 }{} & & \SI{ 1.33e-01 }{} &  \\
    \SI{ 2e-05 }{} & \SI{ 7.44e-03 }{} & 1.06 & \SI{ 6.41e-02 }{} & 1.06 \\
    \SI{ 1e-05 }{} & \SI{ 3.66e-03 }{} & 1.02 & \SI{ 3.15e-02 }{} & 1.02 \\
    \SI{ 5e-06 }{} & \SI{ 1.82e-03 }{} & 1.01 & \SI{ 1.57e-02 }{} & 1.01 \\
    \bottomrule

  \end{tabular}
  \label{tab:test_case_fem}
\end{table}

\begin{table}[p]
  \caption{Relative errors in bubble position $z$ and flow rate $Q$ at
    $t = \SI{0.00144}{\second}$ and estimated convergence orders for
    the links-in-series test case computed with the midpoint method}
  \centering
  \begin{tabular}{c c c c c}

    \toprule
    $\Delta t$ $(\si{\second})$ & $z$-error & $z$-order & $Q$-error & $Q$-order \\
    \midrule
    \SI{ 8e-05 }{} & \SI{ 1.67e-02 }{} & & \SI{ 1.44e-01 }{} &  \\
    \SI{ 4e-05 }{} & \SI{ 4.24e-03 }{} & 1.98 & \SI{ 3.65e-02 }{} & 1.98 \\
    \SI{ 2e-05 }{} & \SI{ 1.08e-03 }{} & 1.97 & \SI{ 9.33e-03 }{} & 1.97 \\
    \SI{ 1e-05 }{} & \SI{ 2.86e-04 }{} & 1.92 & \SI{ 2.46e-03 }{} & 1.92 \\
    \bottomrule

  \end{tabular}
  \label{tab:test_case_mpm}
\end{table}

\begin{table}[p]
  \caption{Relative errors in bubble position $z$ and flow rate $Q$ at
    $t = \SI{0.00144}{\second}$ and estimated convergence orders for
    the links-in-series test case computed with the semi-implicit
    method} \centering
  \begin{tabular}{c c c c c}

    \toprule
    $\Delta t$ $(\si{\second})$ & $z$-error & $z$-order & $Q$-error & $Q$-order \\
    \midrule
    \SI{ 4e-05 }{} & \SI{ 1.39e-02 }{} & & \SI{ 1.18e-01 }{} &  \\
    \SI{ 2e-05 }{} & \SI{ 6.98e-03 }{} & 0.99 & \SI{ 5.97e-02 }{} & 0.98 \\
    \SI{ 1e-05 }{} & \SI{ 3.51e-03 }{} & 0.99 & \SI{ 3.01e-02 }{} & 0.99 \\
    \SI{ 5e-06 }{} & \SI{ 1.76e-03 }{} & 1.00 & \SI{ 1.51e-02 }{} & 1.00 \\
    \bottomrule

  \end{tabular}
  \label{tab:test_case_sim}
\end{table}

We choose the pressure difference to be $\Delta P =
\SI{-3200}{\pascal}$. This value is large enough to overcome the
capillary forces and push the non-wetting bubble through the links. We
therefore expect a flow rate $Q$ that varies in time, but is always
positive.

As measures of the numerical error, we consider both the relative
error in the flow rate $Q$ and the relative error in bubble position
$z$ between the numerical solutions and reference solutions at the end
of the simulation. Time integration is performed from $t =
\SI{0}{\second}$ to $t = \SI{0.00144}{\second}$. To have control over
the time step lengths, we ignore all time step criteria for now and
instead set a constant $\Delta t$ for each simulation.

In Figure \ref{fig:z_Q_test_case}, flow rates are plotted for each of
the time integration methods. Results using a coarse time step,
$\Delta t = \SI{4e-5}{\second}$, and a fine time step, $\Delta t =
\SI{1e-5}{\second}$, are shown along with the reference solution.

For the forward Euler and the semi-implicit method, there is
considerable discrepancy between the numerical and the reference
solution with the coarse time step. The flow rate obtained from
forward Euler lags behind the reference solution, while that from the
semi-implicit method lies ahead of it. This may be expected, however,
since forward Euler at each time step uses current information in the
right hand side evaluation, whereas the semi-implicit method uses a
combination of current and future information. With the fine time
step, there is less difference between the reference and the numerical
solutions. With the more accurate midpoint method, the coarse-stepped
numerical solution lies only marginally ahead of the reference
solution while there is no difference between the fine-stepped
numerical solution and the reference solution at the scale of
representation.

The convergence of the numerical solutions to the reference solution
upon time step refinement is quantified in Table
\ref{tab:test_case_fem}, Table \ref{tab:test_case_mpm} and Table
\ref{tab:test_case_sim}. Herein, the numerical errors and estimated
convergence orders are given for the forward Euler, midpoint and
semi-implicit method, respectively. For all methods considered, the
numerical errors decrease when the time step is refined and do so at
the rate that is expected. The forward Euler and the semi-implicit
method exhibit first-order convergence, while the midpoint method
shows second-order convergence. We note that the errors in both $z$
and $Q$ are similar in magnitude for the forward Euler and the
semi-implicit method. The errors obtained with the midpoint method are
smaller. The difference is one order of magnitude for $\Delta t =
\SI{1e-5}{\second}$.

In summary, we have verified that the presented time integration
methods correctly solve the pore network model equations for the
links-in-series test case and that the numerical errors decrease upon
time step refinement at the rate that is consistent with the expected
order of the methods.

\subsection{Stability tests}
\label{sec:stability_tests}

In this section, we demonstrate that the proposed capillary time step
criterion \eqref{eq:dt_c} stabilizes the forward Euler method and the
midpoint method at low flow rates. We simulated two different cases
and varied $C_\capillary$. Simulations run with low $C_\capillary$
turned out to be free of spurious oscillations, indicating that the
proposed criterion stabilizes the methods, while simulations run with
$C_\capillary$ significantly larger than unity produced oscillations,
indicating that the proposed criterion is not unnecessarily strict.

\begin{figure}[tbp]
  \centering
  \begin{subfigure}[b]{0.48\textwidth}
    \includegraphics[width=\textwidth]{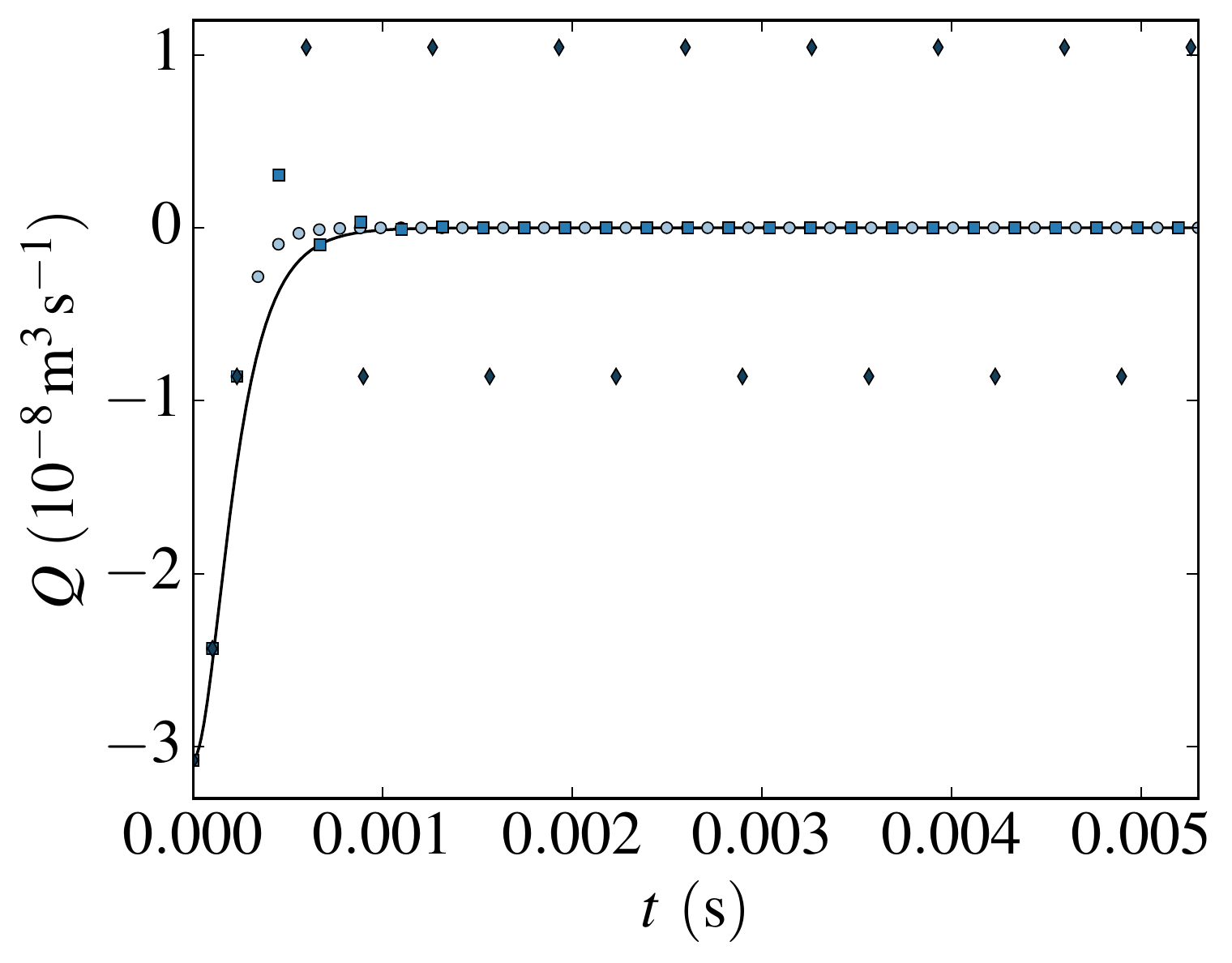}
    \caption{}
    \label{fig:q_vs_t_c_c_fem}
  \end{subfigure}
  ~
  \begin{subfigure}[b]{0.48\textwidth}
    \includegraphics[width=\textwidth]{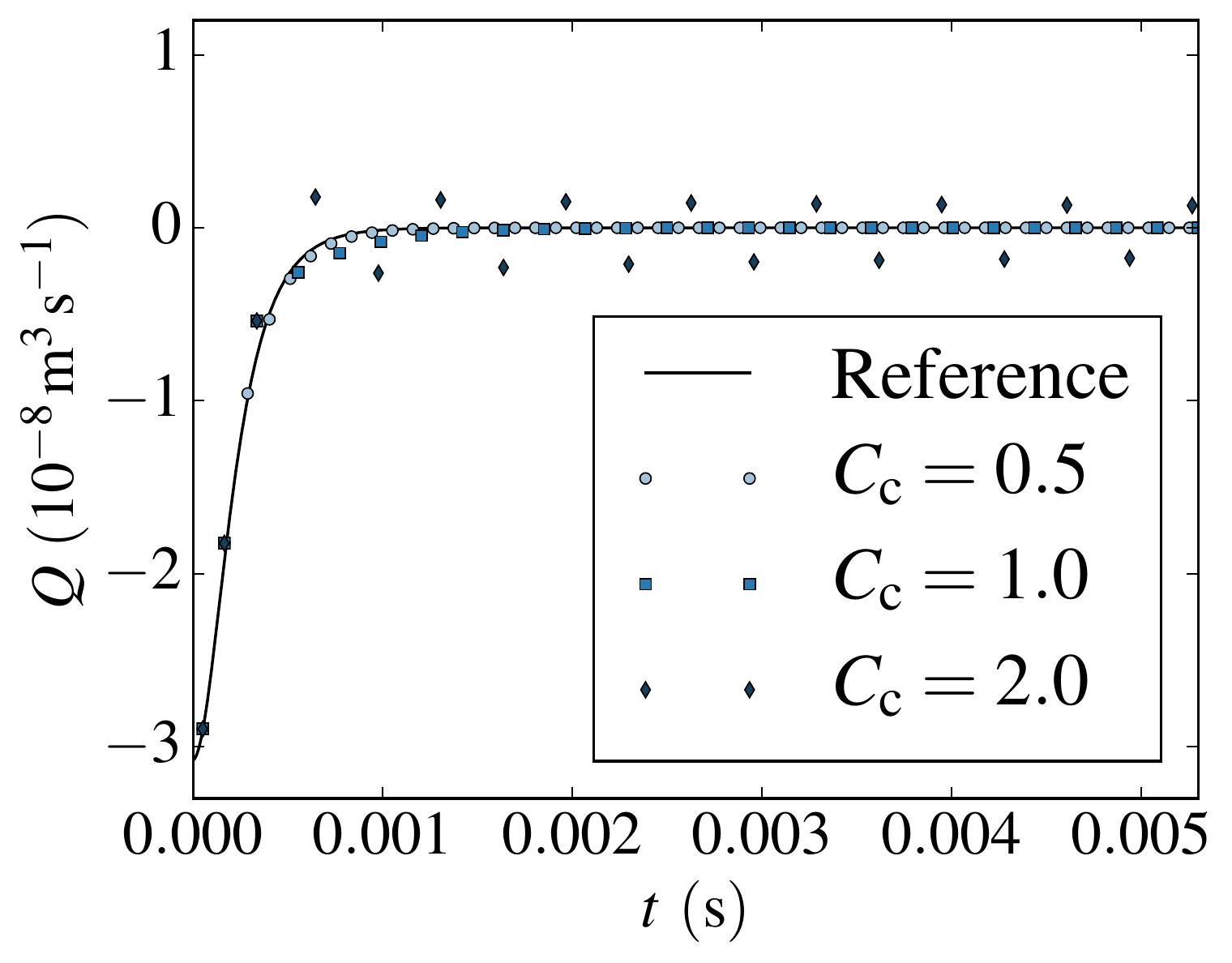}
    \caption{}
    \label{fig:q_vs_t_c_c_mpm}
  \end{subfigure}
  \begin{subfigure}[b]{0.48\textwidth}
    \includegraphics[width=\textwidth]{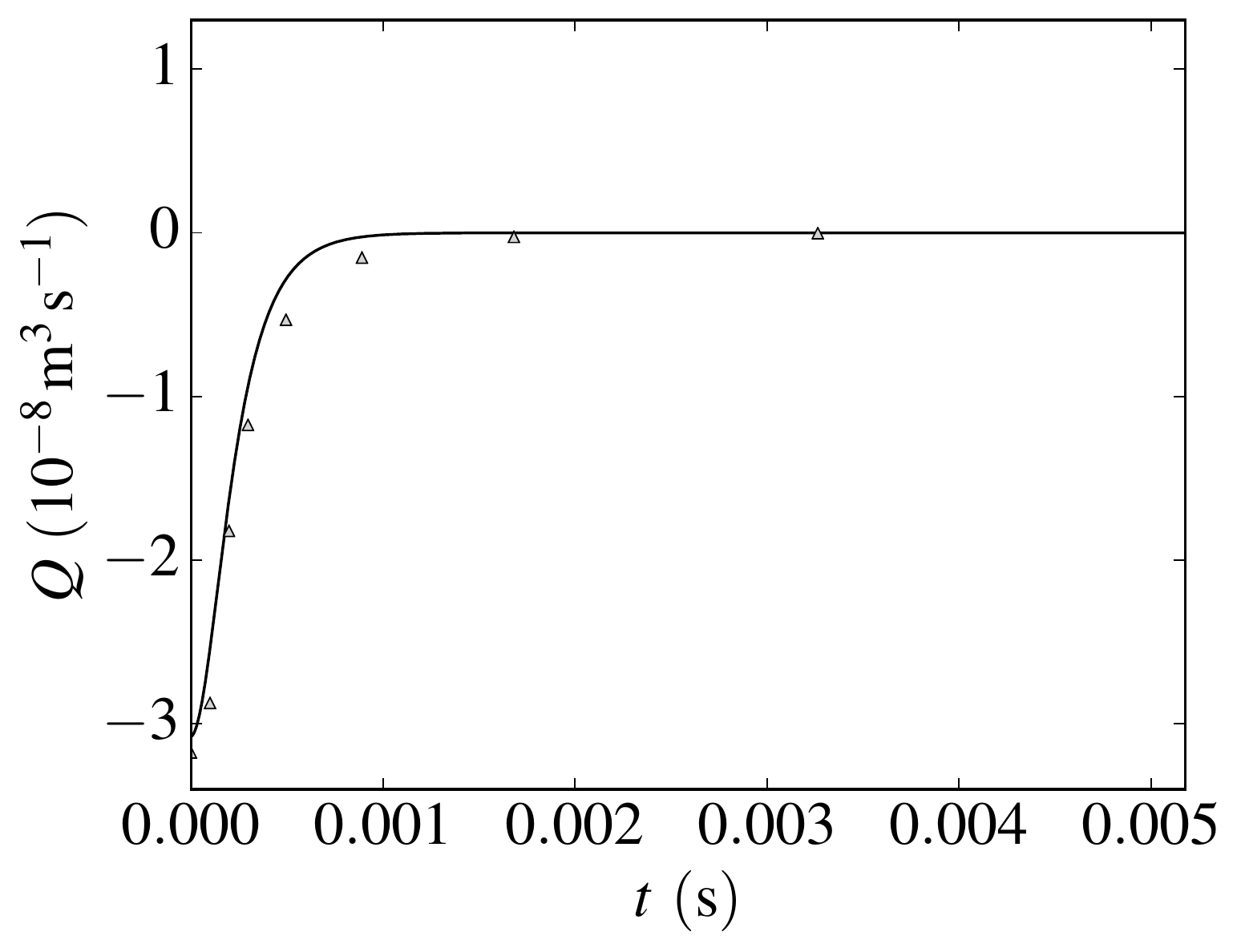}
    \caption{}
    \label{fig:q_vs_t_c_c_sim}
  \end{subfigure}
  \caption{Flow rate plotted against time in the link-in-series test
    case with $\Delta P = \SI{0}{\pascal}$. Results from the forward
    Euler method (a) and the midpoint method (b) are shown for
    different values of $C_\capillary$. Severe numerical instabilities
    arise when $C_\capillary = 2.0$.  Results from the semi-implicit
    method are shown are shown in (c).  These are stable, even if the
    capillary time step criterion is not used. The solid black line
    represents a reference solution.}
\end{figure}

First, consider the links-in-series test case with $\Delta P =
\SI{0}{\pascal}$. With no applied pressure difference, the flow is
driven purely by the imbalance of capillary forces on the non-wetting
bubble. Therefore, there should only be flow initially and the bubble
should eventually reach an equilibrium position where both interfaces
experience the same capillary force and the flow rate is
zero. Simulations were run with $C_\advective = 0.1$ and
$C_\capillary$ equal to 2.0, 1.0 and 0.5. Results from forward Euler
are shown in Figure~\ref{fig:q_vs_t_c_c_fem} and results from the
midpoint method are shown in Figure~\ref{fig:q_vs_t_c_c_mpm}. In both
figures, the reference solution is also shown.

\begin{figure}[tbp]
  \centering
  \begin{subfigure}[b]{0.50\textwidth}
    \includegraphics[width=\textwidth]{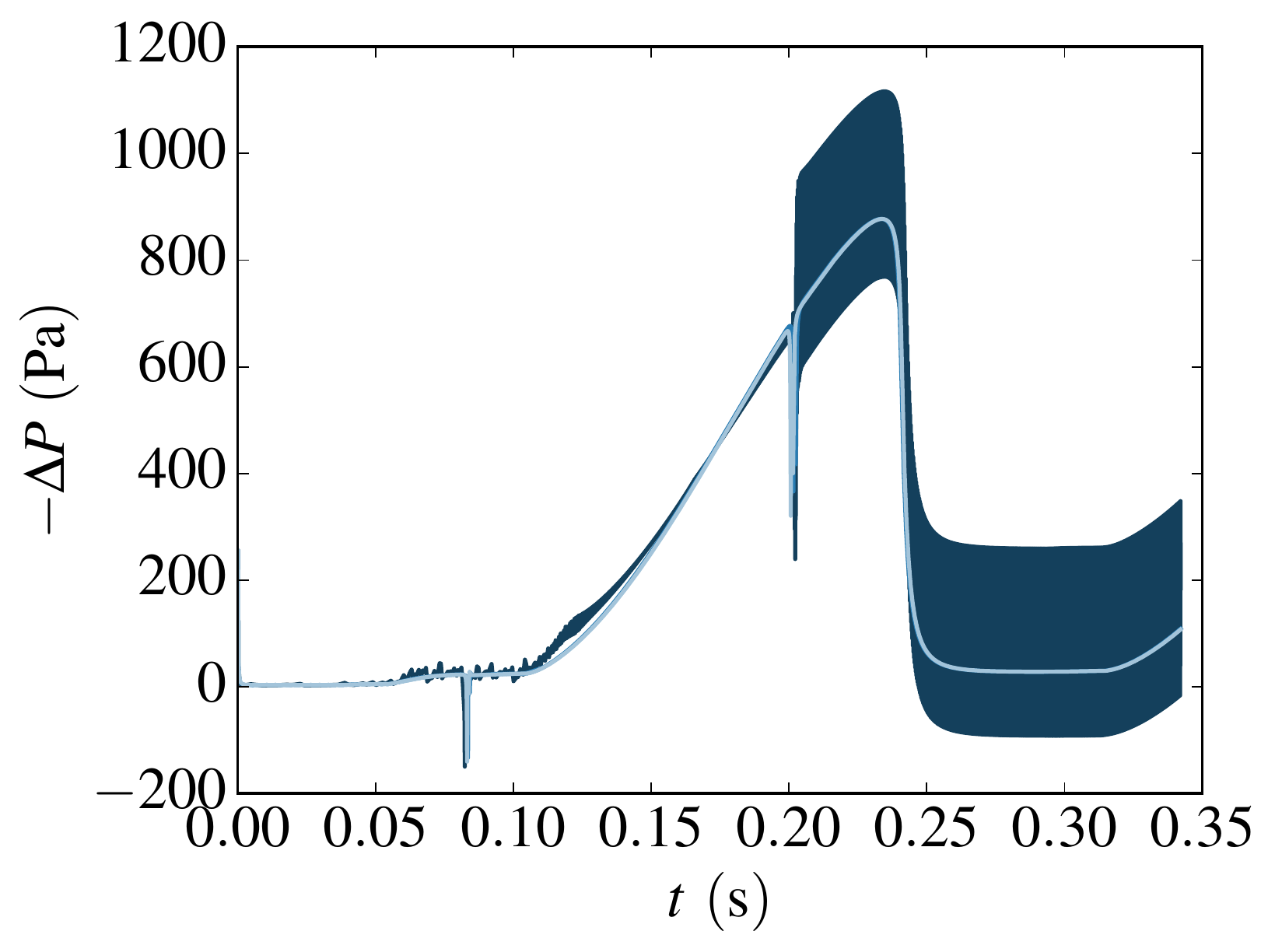}
    \caption{}
    \label{fig:Haines_dP_vs_t}
  \end{subfigure}
  ~
  \begin{subfigure}[b]{0.46\textwidth}
    \includegraphics[width=\textwidth]{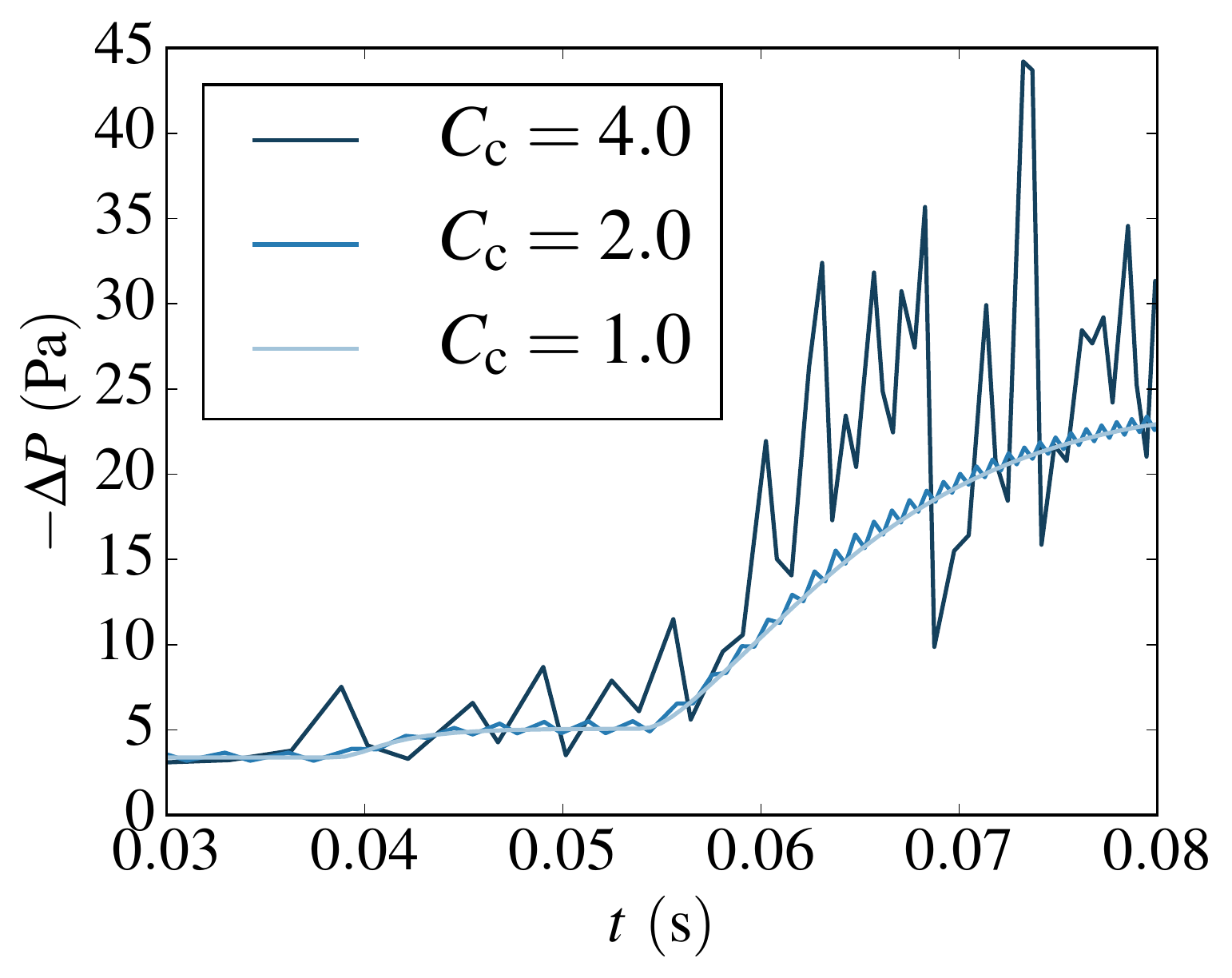}
    \caption{}
    \label{fig:Haines_dP_vs_t_cropped}
  \end{subfigure}
  \caption{Pressure difference required to drive the flow in the
    Haines jump case at a rate of $Q =
    \SI{e-9}{\meter\cubed\per\second}$, corresponding to $\ca
    =\SI{1.2e-5}{}$. In (b), the results from (a) are shown in greater
    detail. Results are computed with the forward Euler method for
    different values of the capillary time step restriction parameter
    $C_\capillary$. Numerical instabilities are seen to occur for
    $C_\capillary > 1$.}
\end{figure}

The forward Euler results are stable and qualitatively similar to the
reference solution with $C_\capillary = 0.5$. With $C_\capillary =
1.0$, there are some oscillations initially that are dampened and
eventually vanish. From comparison with the reference solution, it is
clear that such oscillations have no origin in the model equations and
are artifacts of the numerical method. With $C_\capillary = 2.0$, the
oscillations are severe and do not appear to be dampened by the
method. Instead the non-wetting bubble keeps oscillating around its
equilibrium position in a manner that is clearly unphysical.

The results from the midpoint method in
Figure~\ref{fig:q_vs_t_c_c_mpm} follow a qualitatively similar trend
as those from forward Euler with regard to stability. Results computed
with $C_\capillary = 0.5$ are stable and results with $C_\capillary =
2.0$ exhibit severe oscillations. Still, the results from the midpoint
method lie much closer to the reference solution than the results from
the forward Euler method, as we would expect since the midpoint method
is second-order. Both methods are, however, unstable with
$C_\capillary = 2.0$, indicating that the while the midpoint method
has improved accuracy over forward Euler, it is unable to take
significantly longer time steps without introducing oscillations. This
is consistent with the analysis in \ref{sec:capillary_time_step},
since the two methods have identical stability regions in real space.

Next, consider the Haines jump case with $Q =
\SI{e-9}{\meter\cubed\per\second}$, corresponding to $\ca =
\SI{1.2e-5}{}$. This case was run using the forward Euler method,
$C_\advective = 0.1$ and three different values of $C_\capillary$,
equal to 4.0, 2.0 and 1.0. The required pressure difference to drive
the flow at the specified rate is shown in Figure
\ref{fig:Haines_dP_vs_t}. Figure \ref{fig:Haines_dP_vs_t_cropped}
shows the pressure from Figure \ref{fig:Haines_dP_vs_t} in greater
detail.

\begin{figure}[tbp]
  \centering
  \begin{subfigure}[b]{0.48\textwidth}
    \includegraphics[width=\textwidth]{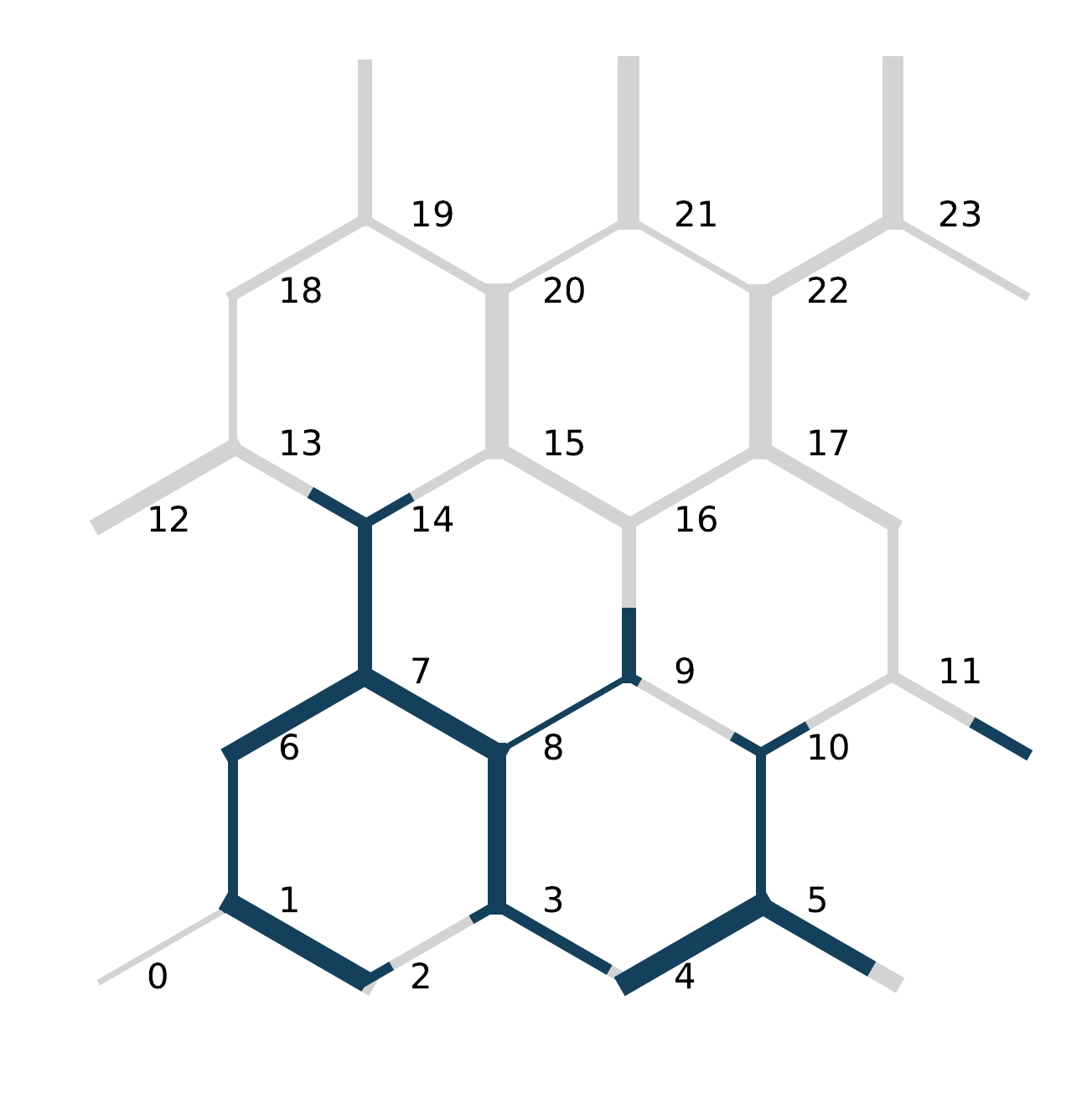}
    \caption{}
  \end{subfigure}
  ~
  \begin{subfigure}[b]{0.48\textwidth}
    \includegraphics[width=\textwidth]{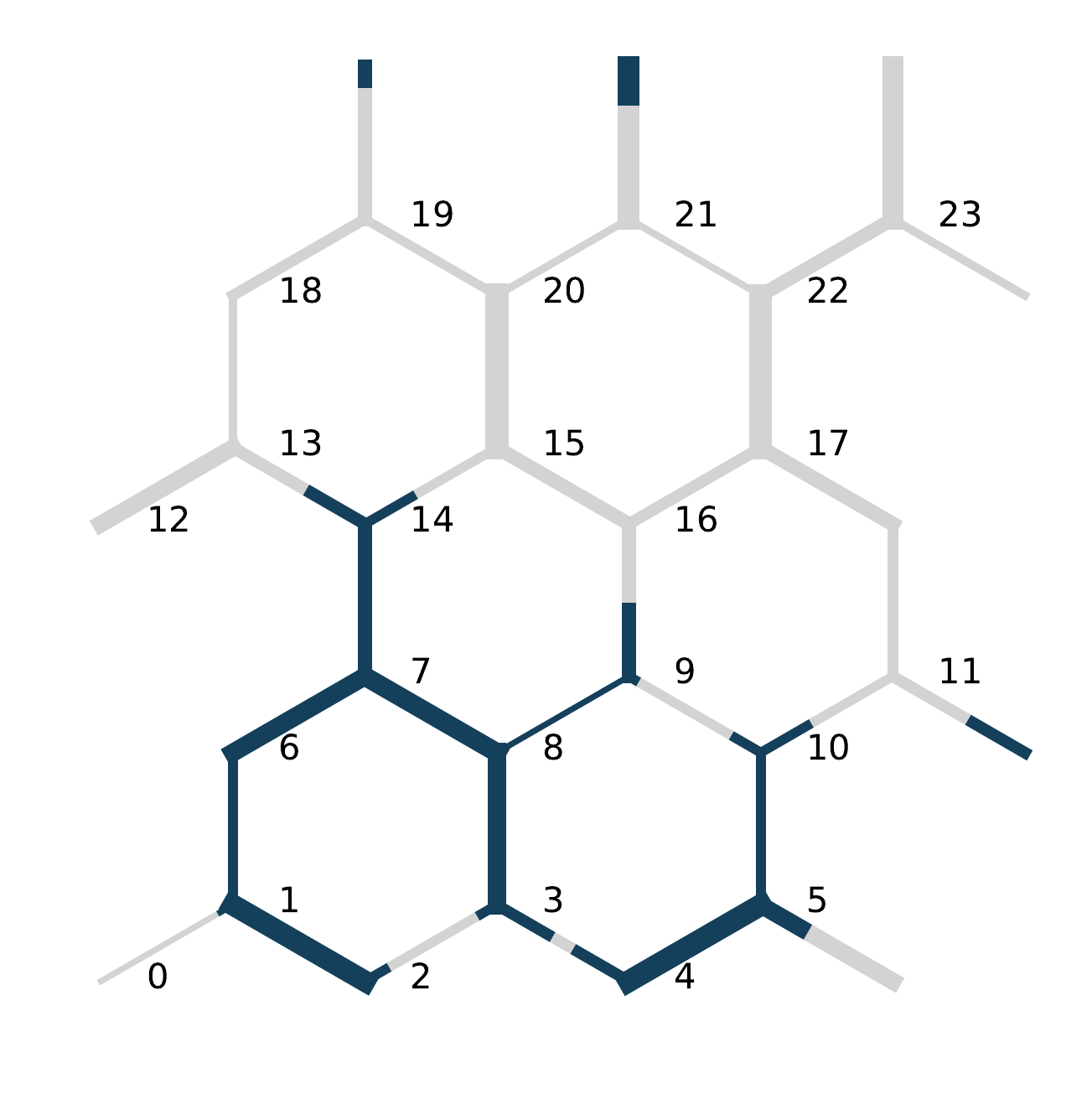}
    \caption{}
  \end{subfigure}
  \caption{Fluid distribution in the Haines jump case, (a) at $t =
    \SI{0.19}{\second}$ and (b) at $t = \SI{0.21}{\second}$, before
    and after the fluid redistribution event at $t \approx
    \SI{0.20}{\second}$. The link radii are not drawn to scale with
    the link lengths. Node indices are indicated in black.}
  \label{fig:redistribution_fluid}
\end{figure}

\begin{figure}[tbp]
  \centering
  \begin{subfigure}[b]{0.48\textwidth}
    \includegraphics[width=\textwidth]{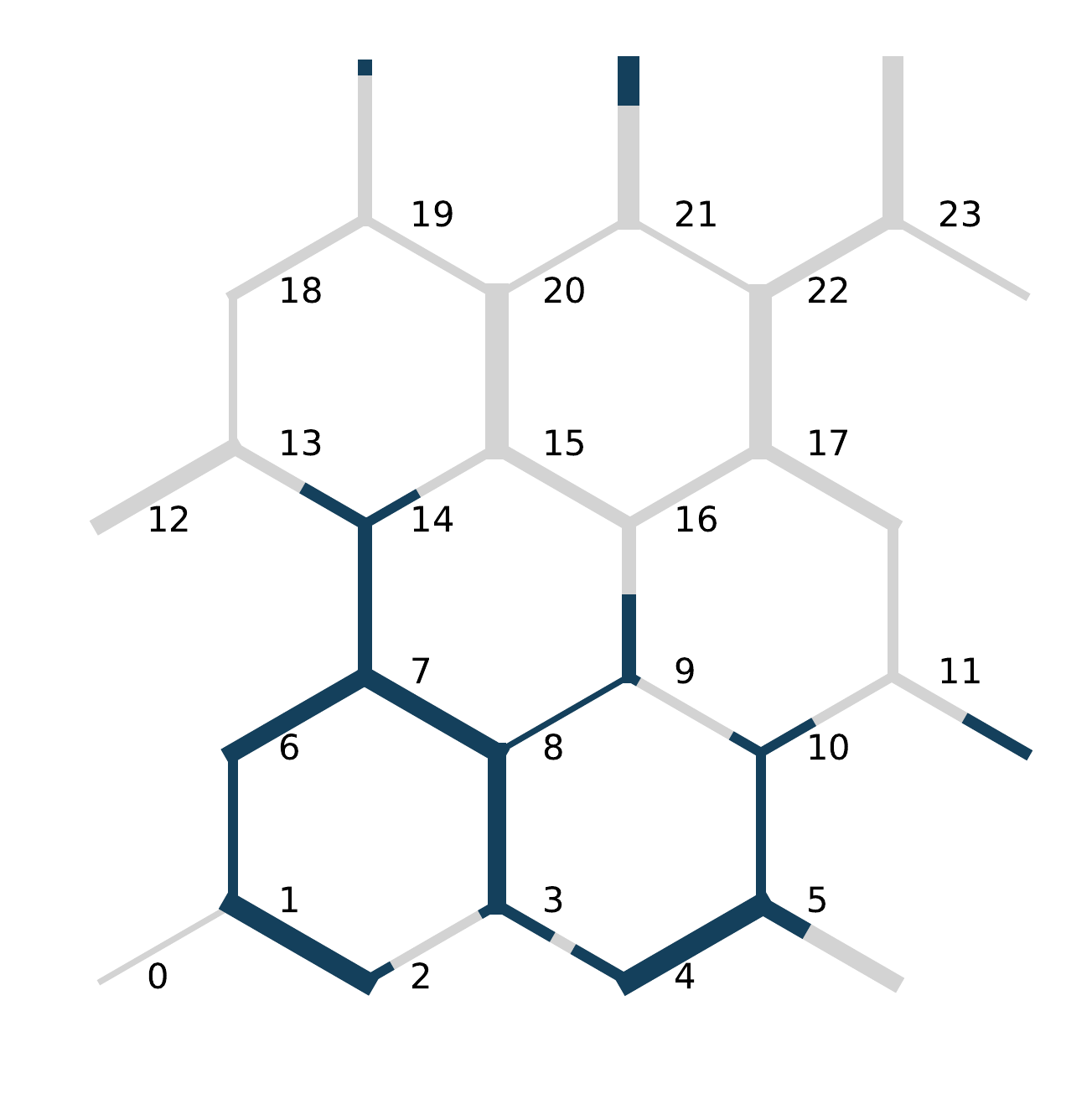}
    \caption{}
  \end{subfigure}
  ~
  \begin{subfigure}[b]{0.48\textwidth}
    \includegraphics[width=\textwidth]{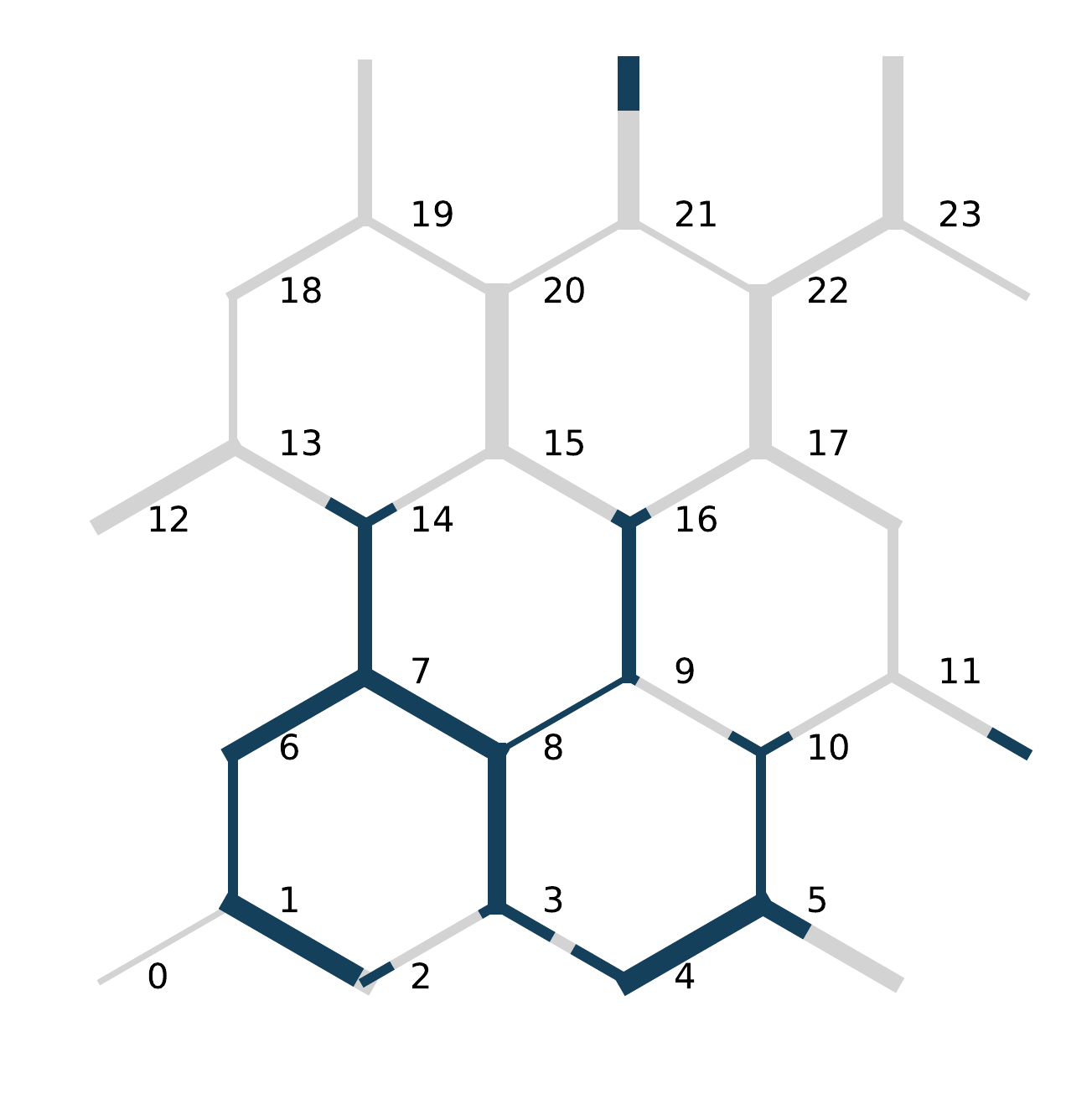}
    \caption{}
  \end{subfigure}
  \caption{Fluid distribution in the Haines jump case (a) at $t =
    \SI{0.23}{\second}$ and (b) at $t = \SI{0.27}{\second}$, before
    and after the Haines jump. During the jump, non-wetting fluid
    breaks-through the link connecting nodes 9 and 16 and invades node
    16. Also, non-wetting fluid in other links at the downstream end
    of the moving ganglion retracts. This is seen e.g.\ in the links
    downstream of nodes 10 and 14. The link radii are not drawn to
    scale with the link lengths. Node indices are indicated in black.}
  \label{fig:haines_fluid}
\end{figure}

For all three values of $C_\capillary$, the main qualitative features
of the flow are captured. We observe short transient pressure drops at
$t \approx \SI{0.08}{\second}$ and $t \approx
\SI{0.20}{\second}$. These correspond to fluid redistribution events
on the upstream side of the non-wetting ganglion, where the fluid
rearranges itself to a more stable configuration with little change
to the interface positions on the downstream side. The event at $t
\approx \SI{0.20}{\second}$ is illustrated in Figure
\ref{fig:redistribution_fluid}. The fluid redistribution is driven by
capillary forces and less external pressure is therefore required to
drive the flow during these events.

We also observe the slow pressure build-up from $t \approx
\SI{0.10}{\second}$ to $t \approx \SI{0.23}{\second}$, when the
driving pressure becomes large enough to overcome the capillary forces
and cause break-through of non-wetting fluid in the link connecting
nodes 9 and 16, and we observe the subsequent Haines jump. The fluid
configurations before and after the Haines jump are shown in Figure
\ref{fig:haines_fluid}. Notice also that non-wetting fluid at the
downstream end of the moving ganglion retracts during the Haines jump
in links near to where the break-through occurs. This is seen e.g.\ in
the links downstream of nodes 10 and 14. That such local imbibition
occurs near the drained pore is in agreement with the observations of
\citet{Armstrong2013}, and shows that the model is able to capture the
non-local nature of pore drainage events in a numerically stable
manner when the new numerical methods are used.

As in the links-in-series case, the solution exhibits oscillations for
the values of $C_\capillary$ that are larger than unity. With
$C_\capillary = 1.0$, the results are free from oscillations and
appear stable. This indicates that the stability criterion
\eqref{eq:dt_c} is valid and not unnecessarily strict also for a
network configuration that is much more complex than links in series.

Both the links-in-series case and the Haines jump case were simulated
with the semi-implicit method and produced stable results with the
advective time step criterion \eqref{eq:dt_a} only. The results from
the links-in-series test case are shown in
Figure~\ref{fig:q_vs_t_c_c_sim}. For brevity, the results from the
Haines jump case are omitted here. The reader is referred to
Figure~\ref{fig:Haines_dp_vs_t} in Section \ref{sec:performance},
where stable results are shown for a lower flow rate.

To summarize, both the forward Euler and midpoint methods produce
stable results for the cases considered when the capillary time step
criterion \eqref{eq:dt_c} is used in addition to \eqref{eq:dt_a} to
select the time step lengths. By running simulations with different
$C_\capillary$, we have observed a transition from stable to unstable
results for values of $C_\capillary$ near 1, in order of magnitude. In
the Haines jump case, all methods presented are able to capture both
the fast capillary-driven fluid redistribution events, and the slow
pressure build-up before a Haines jump.

\section{Performance analysis}
\label{sec:performance}

In this section, we analyze and compare the performance of the time
integration methods. In doing so, we consider the number of time steps
and the wall clock time required to perform stable simulations of the
Haines jump case with each of the methods at different specified flow
rates $Q$. The flow rates simulated were
$\SI{e-7}{\meter\cubed\per\second}$,
$\SI{e-8}{\meter\cubed\per\second}$,
$\SI{e-9}{\meter\cubed\per\second}$,
$\SI{e-10}{\meter\cubed\per\second}$,
$\SI{e-11}{\meter\cubed\per\second}$ and
$\SI{e-12}{\meter\cubed\per\second}$. The accuracy of the methods was
studied Section \ref{sec:convergence_tests}, and will not be part of
the performance analysis. Instead, stable simulations are considered
sufficiently accurate.

\begin{figure}[tbp]
    \centering
    \begin{subfigure}[b]{0.6\textwidth}
        \includegraphics[width=\textwidth]{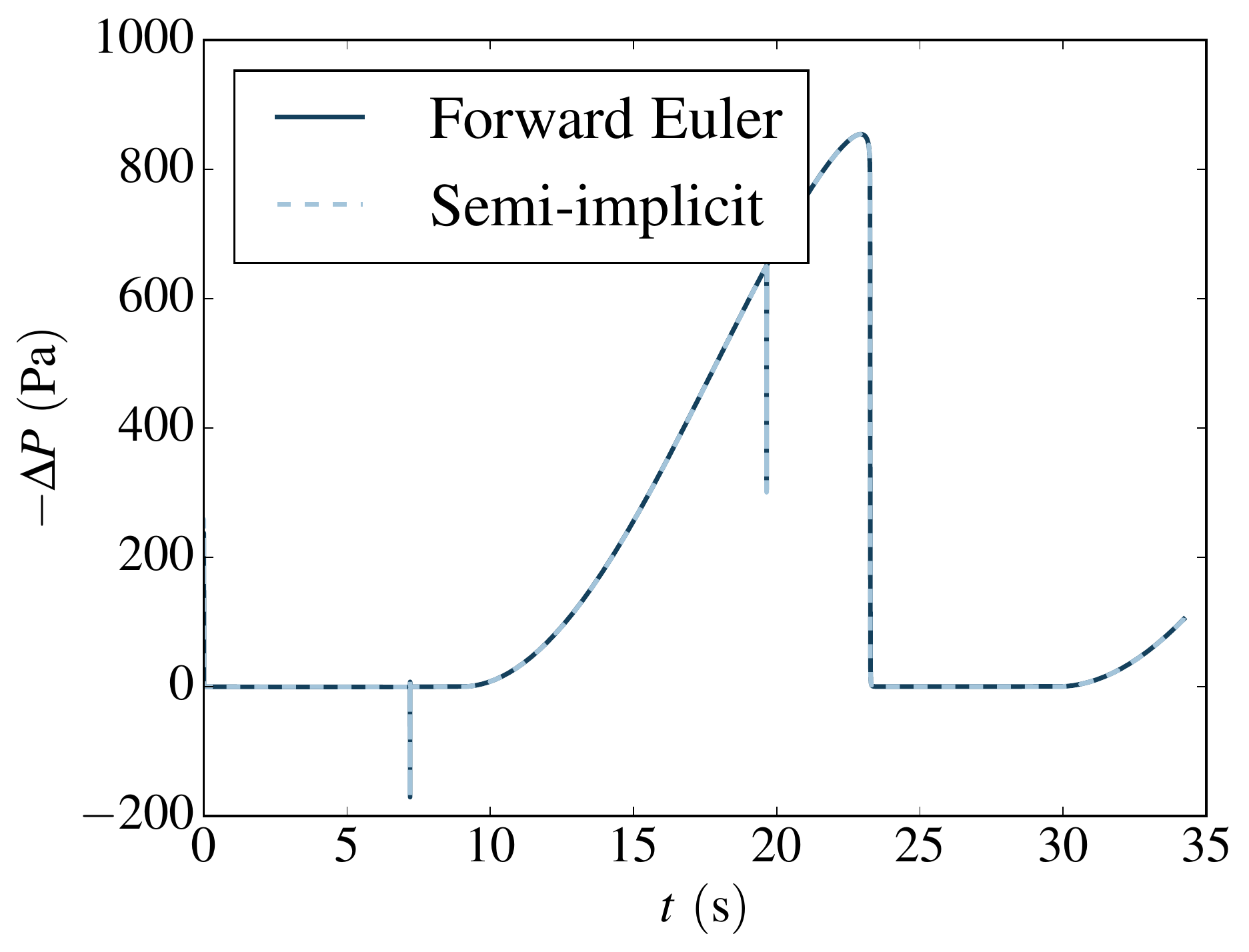}
        \caption{}
        \label{fig:Haines_dp_vs_t}
    \end{subfigure}
    \begin{subfigure}[b]{0.6\textwidth}
        \includegraphics[width=\textwidth]{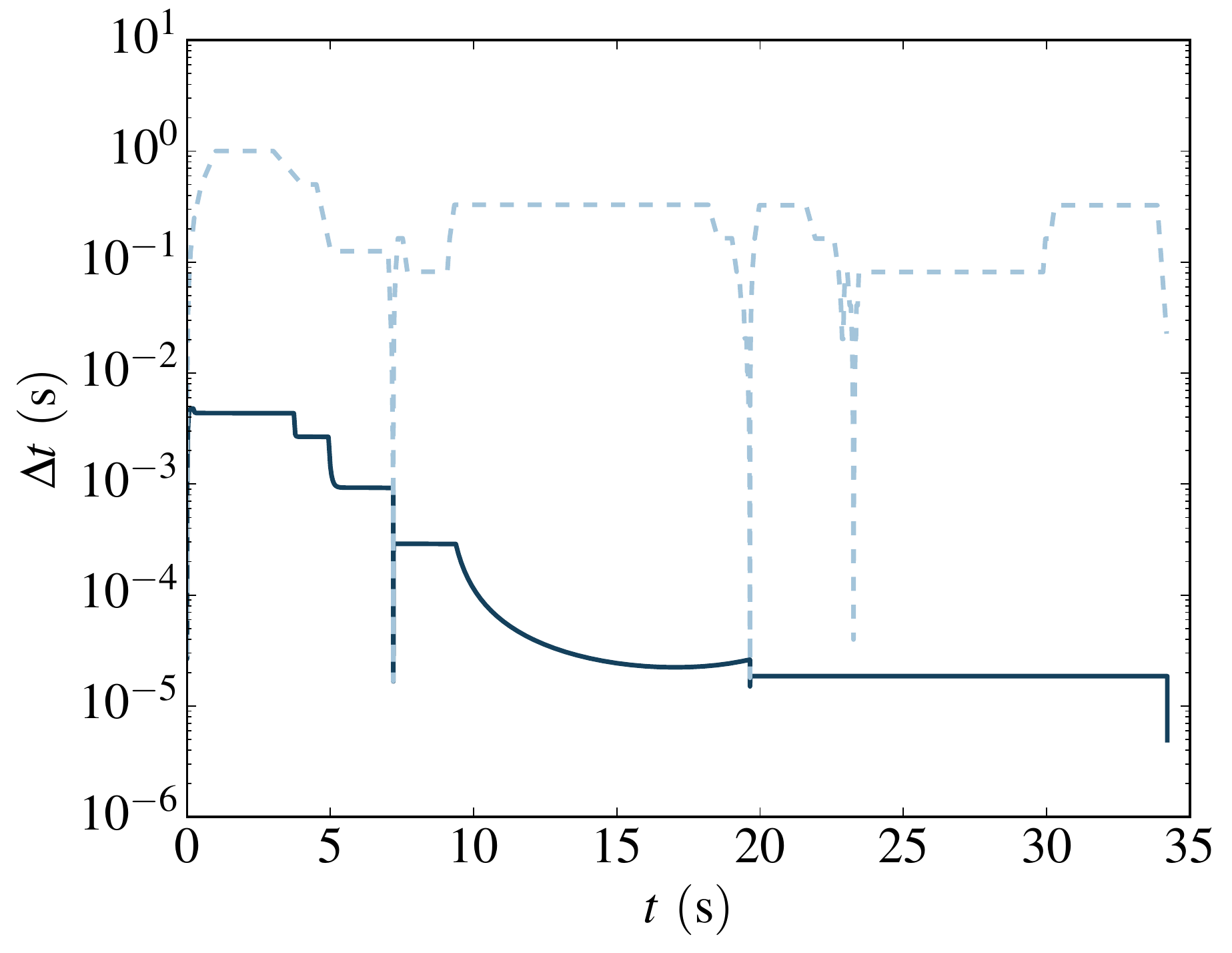}
        \caption{}
        \label{fig:Haines_dt_vs_t}
    \end{subfigure}
    \caption{(a) Pressure difference required to drive the flow at
      $Q=\SI{e-11}{\meter\cubed\per\second}$, corresponding to $\ca =
      \SI{1.2e-7}{}$, in the Haines jump case. Results are plotted for
      the forward Euler method (solid dark blue) and the semi-implicit
      method (dashed light blue). These lines coincide at the scale of
      representation. The time steps lengths used by each method are
      plotted in (b).}
    \label{fig:Haines_dp_vs_t_dt_vs_t}
\end{figure}

First, we look more closely at the results for
$Q=\SI{e-11}{\meter\cubed\per\second}$, corresponding to $\ca =
\SI{1.2e-7}{}$. The pressure difference required to drive the flow is
shown in Figure \ref{fig:Haines_dp_vs_t}, and the time step lengths
used are shown in Figure \ref{fig:Haines_dt_vs_t}. From the latter
Figure, we see that the semi-implicit method is able to take longer
time steps than forward Euler for most of the simulation. During the
pressure build-up phase, the difference is four orders of
magnitude. During the fast capillary-driven fluid redistribution
events, however, the length of the semi-implicit time steps drop to
the level of those used by forward Euler. This is because we here have
relatively large flow rates in some links, even though $Q$ is low, and
the advective time step criterion \eqref{eq:dt_a} becomes limiting for
both the semi-implicit method and forward Euler.

It was mentioned by \citet{Armstrong2013} that any accurate numerical
simulation on the pore scale must have a time resolution fine enough
to capture the fast events. The semi-implicit method accomplishes this
by providing a highly dynamic time resolution, which is refined during
the fast events. The method is therefore able to resolve these events,
while time resolution can be coarsened when flow is governed by the
slow externally applied flow rate, saving computational effort.

The time duration of the Haines jump pressure drops for all except the
two largest externally applied flow rates were around
$\SI{10}{\milli\second}$. This is in qualitative agreement with the
results presented by \citet{Armstrong2013}. They found that, for their
investigated range of parameters, pores were drained on the
millisecond time scale regardless of externally applied flow
rate. However, we stress that although we consider the same fluids, the
pore network used here was approximately one order of magnitude larger
in the linear dimensions than that of \cite{Armstrong2013}.

\begin{figure}
  \centering
  \begin{subfigure}[b]{0.48\textwidth}
    \includegraphics[width=\textwidth]{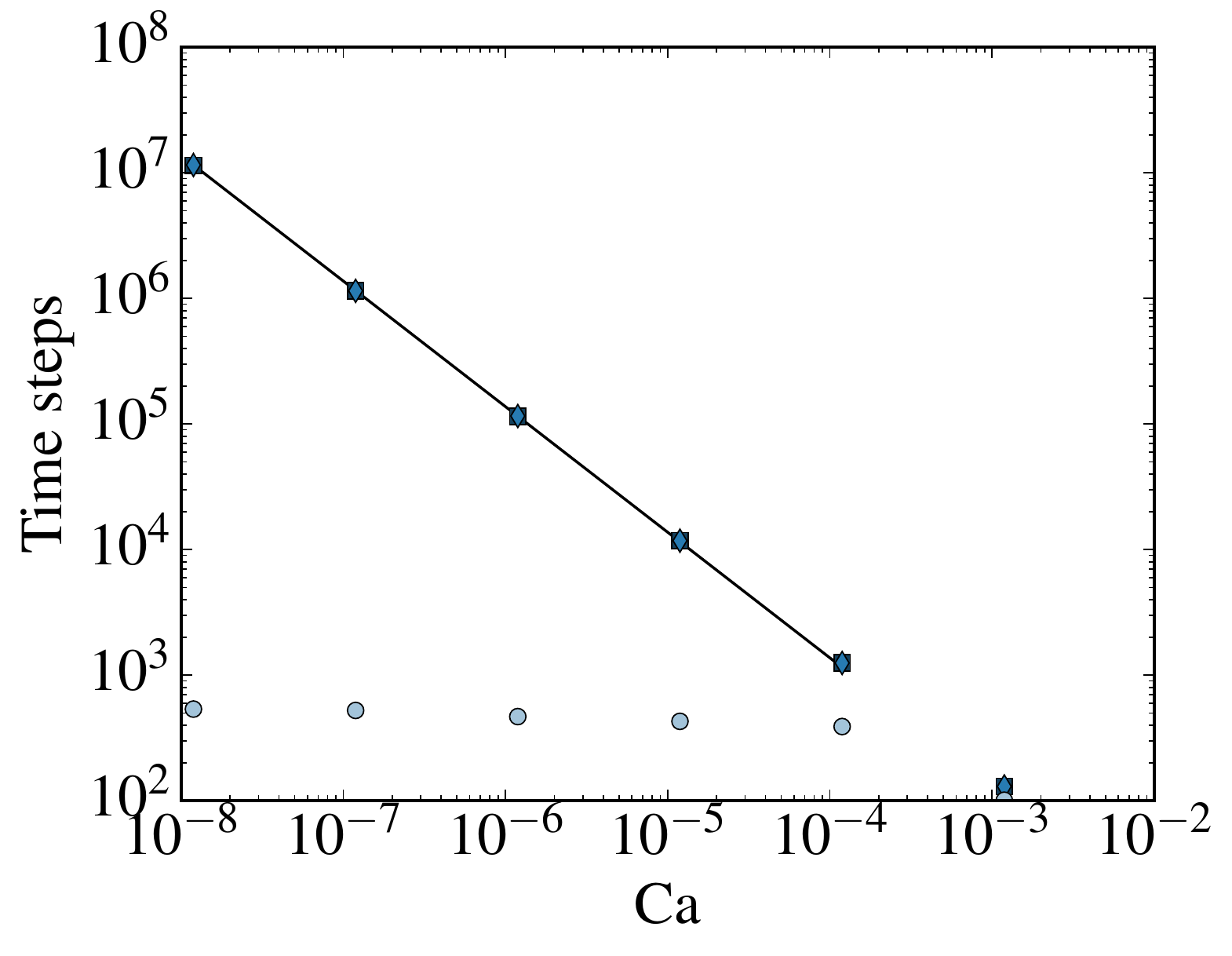}
    \caption{}
    \label{fig:Haines_nsteps_vs_Ca}
  \end{subfigure}
  \begin{subfigure}[b]{0.492\textwidth}
    \includegraphics[width=\textwidth]{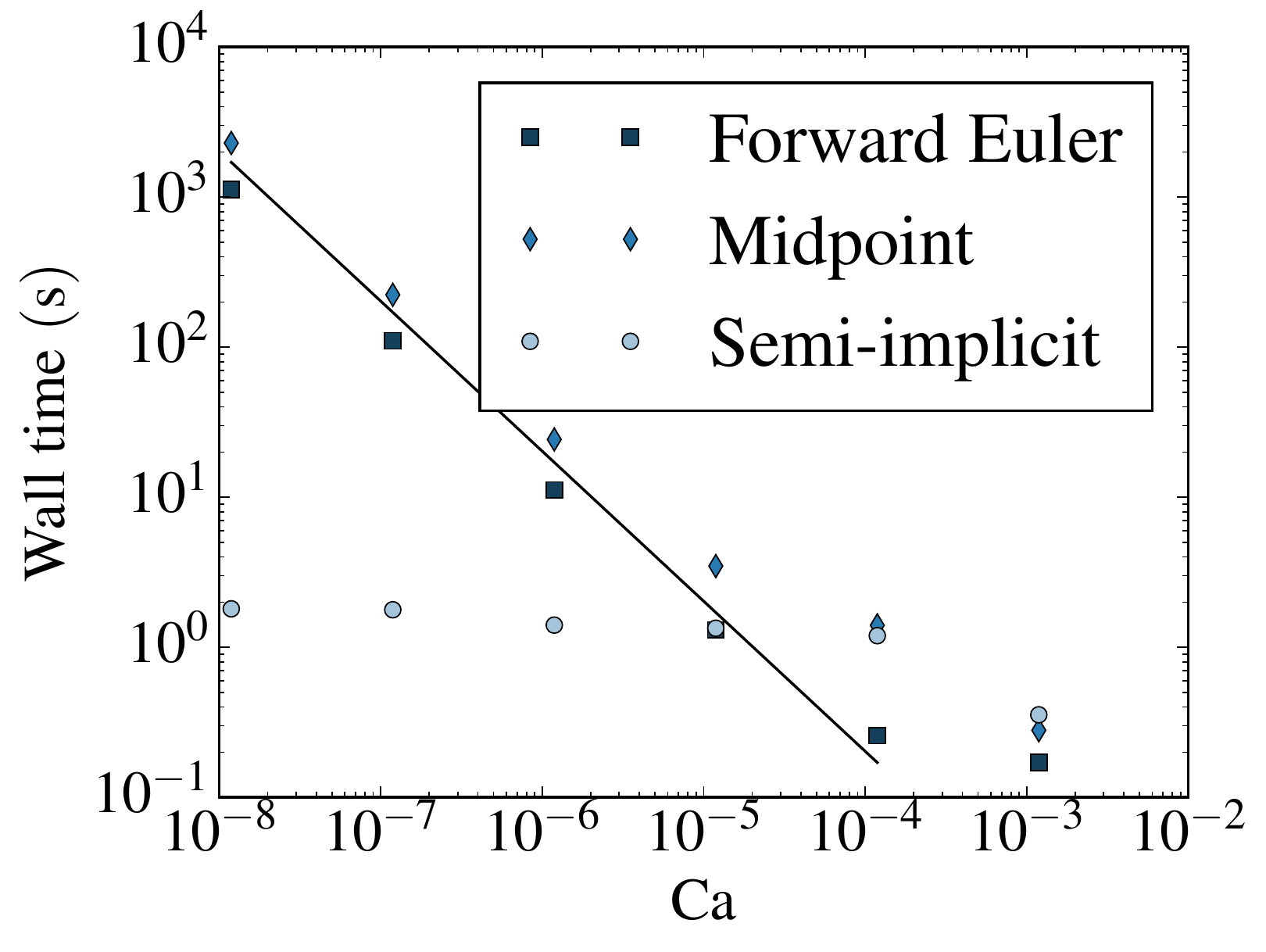}
    \caption{}
    \label{fig:Haines_wall_time_vs_Ca}
  \end{subfigure}
  \caption{(a) Number of time steps and (b) wall clock time required
    to simulate the Haines jump case at at different specified flow
    rates. In each simulation, the same volume of fluid
    (\SI{5}{\percent} of the pore volume) flows through a
    network. Results from the forward Euler method (squares), the
    midpoint method (diamonds) and the semi-implicit method (circles)
    are shown. In both (a) and (b), the black lines are inversely
    proportional to $\ca$.}
\end{figure}

The number of time steps and wall clock time required to simulate the
Haines jump case at different specified flow rates $Q$ are shown in
Figure \ref{fig:Haines_nsteps_vs_Ca} and Figure
\ref{fig:Haines_wall_time_vs_Ca}, respectively. 

For the explicit methods, both the number of time steps and the wall
time are proportional to $\ca^{-1}$ at low capillary numbers. This is
because the capillary time step criterion \eqref{eq:dt_c} dictates the
time step at low capillary numbers (except during fast fluid
redistribution events). The criterion depends on the fluid
configuration, while it is independent of the flow rate. At low enough
flow rates, the system will pass through roughly the same fluid
configurations during the simulation, regardless of the applied
$Q$. The speed at which the system passes through these
configurations, however, will be inversely proportional to $Q$ and
therefore, so will the required wall time and number of time steps. As
the forward Euler and the midpoint method are subject to the same time
step criteria, these require roughly the same number of time steps at
all considered flow rates. However, since the midpoint method is a
two-step method, the wall time it requires is longer and approaches twice
that required by the forward Euler for long wall times.

For the semi-implicit method, on the other hand, the number of time
steps required to do the simulation becomes effectively independent of
the specified flow rate at capillary numbers smaller than
approximately $\SI{e-4}{}$. The result is that low-capillary number
simulations can be done much more efficiently than with the explicit
methods, in terms of wall time required to perform stable
simulations. This is seen in Figure
\ref{fig:Haines_wall_time_vs_Ca}. At $\ca \sim \SI{e-5}{}$, the
computational time needed by all three methods are similar in
magnitude. The relative benefit of using the semi-implicit method
increases at lower capillary numbers. For the lowest capillary number
considered, the difference in wall time between the explicit methods
and the semi-implicit is three orders of magnitude.



The increased efficiency of the semi-implicit method over explicit
methods at low capillary numbers means that one can use the
semi-implicit method to perform simulations in the low capillary
number regime that are unfeasible with explicit methods. Thus, the
range of capillary numbers for which the pore network model is a
tractable modeling alternative is extended to much lower capillary
numbers. This includes e.g.\ simulations of water flow in fuel cell
gas diffusion layers, where capillary numbers are can be \SI{e-8}{}
\cite{Sinha2007}.

\begin{figure}[tbp]
  \centering
  \includegraphics[width=0.48\textwidth]{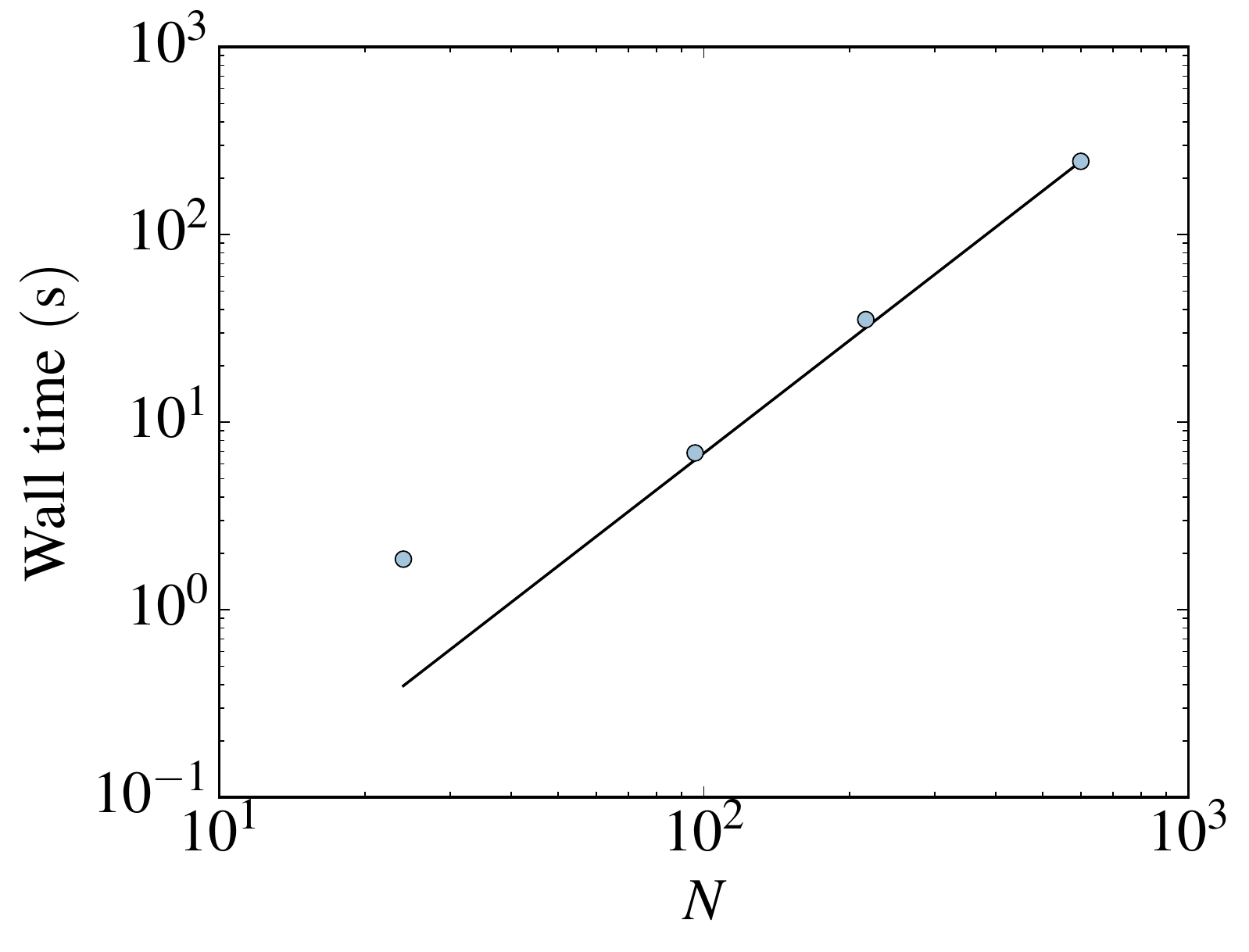}
  \caption{Wall clock time time required to simulate the Haines jump
    case with the semi-implicit method for different network
    sizes. All simulations were run at $\ca \sim \SI{e-7}{}$ and $N$
    denotes the number of nodes in the network. The wall time is seen
    to increase proportionally to $N^2$ for the three largest
    networks.}
  \label{fig:Haines_wall_time_vs_N}
\end{figure}

Finally, to study the effect of an increase in network size on the
wall time required by the semi-implicit method, the Haines jump case
was run on three scaled-up versions of the network with $N=24$ nodes
considered so far, illustrated in Figure
\ref{fig:network_initial}. All simulations were run at $\ca \sim
\SI{e-7}{}$. In Figure \ref{fig:Haines_wall_time_vs_N} the wall clock
time time required is plotted against the number of nodes $N$ for the
different networks.  The wall time is seen to increase proportionally
to $N^2$.

\section{Conclusion}
\label{sec:conclusion}

We have studied three different time integration methods for a pore
network model for immiscible two-phase flow in porous media. Two
explicit methods, the forward Euler and midpoint methods, and a new
semi-implicit method were considered. The explicit methods have been
presented and used in other works
\cite{Aker1998,Knudsen2002,Sinha2012}, and were reviewed here for
completeness. The semi-implicit method was presented here for the
first time, and therefore in detail.

The explicit methods have previously suffered from numerical
instabilities at low capillary numbers. Here, a new time-step
criterion was suggested in order to stabilize them and numerical
experiments were performed demonstrating that stabilization was
achieved.

It was verified that all three methods converged to a reference
solution to a selected test case upon time step refinement. The
forward Euler and semi-implicit methods exhibited first-order
convergence and the midpoint method showed second-order convergence.

Simulations of a single Haines jump were performed. These showed that
the all three methods were able to resolve both pressure build-up
events and fluid redistribution events, including interfacial
retraction after a Haines jump, which may occur at vastly different
time scales when capillary numbers are low. The results from the
Haines jump case were consistent with experimental observations made
by \citet{Armstrong2013}. Fluid redistribution events cannot be
properly captured when using solution methods that have previously
been used at low capillary numbers that e.g.\ do not allow backflow
\cite{Medici2010}.

A performance analysis revealed that the semi-implicit method was able
to perform stable simulations with much less computational effort than
the explicit methods at low capillary numbers. For the case
considered, the computational time needed was approximately the same
for all three methods at $\ca \sim \SI{e-5}{}$. At lower capillary
numbers, the computational time needed by the explicit methods
increased inversely proportional to the capillary number, while the
time needed by the semi-implicit method was effectively constant. At
$\ca \sim \SI{e-8}{}$, the computational time needed by the
semi-implicit methods was therefore three orders of magnitude smaller
than those needed by the explicit methods.

The superior efficiency of the new semi-implicit method over the
explicit methods at low capillary numbers enables simulations in this
regime that are unfeasible with explicit methods. Thus, the range of
capillary numbers for which the pore network model is a tractable
modeling alternative is extended to much lower capillary
numbers. This includes e.g.\ simulations of water flow in fuel cell
gas diffusion layers, where capillary numbers are can be \SI{e-8}{}
\cite{Sinha2007}.

In summary, use of Aker-type pore network models were previously
restricted to relatively high capillary numbers due to numerical
instabilities in the explicit methods used to solve them. With the new
time step criterion presented here, these stability problems are
removed. However, simulations at low capillary numbers still take a
long time and the computational time needed increases inversely
proportional to the capillary number. This problem is solved by the
new semi-implicit method. With this method, the computational time
needed becomes effectively independent of the capillary number, when
capillary numbers are low.

\section*{Author contributions}
To pursue low capillary number simulations with Aker-type pore network
models was proposed by SK, MG and AH in collaboration. MV developed
the particular variation of the pore network model used. MG developed
the new numerical methods and performed the simulations. MG wrote the
manuscript, aided by comments and suggestions from MV, SK and AH.

\section*{Acknowledgments}
The authors would like to thank Dick Bedeaux, Santanu Sinha and Knut
J{\o}rgen M{\aa}l{\o}y for fruitful discussions. Special thanks are
also given to Jon Pharoah for inspiring discussions and
comments. This work was partly supported by the Research Council of Norway
through its Centres of Excellence funding scheme, project number
262644.

\appendix

\section{Capillary time step criterion}
\label{sec:capillary_time_step}

In this section, we identify the cause of the numerical instabilities
that explicit methods suffer from at low flow rates when the capillary
time step criterion \eqref{eq:dt_c} is not obeyed. We also derive this
criterion. The contents of this section are not intended to constitute
a formal proof of the stability of the presented time integration
methods. The results derived herein are based on a linearized
approximation of the pore network model. Although the application of
results from a linearized analysis to general cases is somewhat
simplistic, it is useful for highlighting key difficulties, see
e.g.\ \citep{Suli2006}~pp.\ 347., and for deriving results that can be found to
work in practice. For evidence of the actual stability of the time
integration methods, it is therefore referred to the numerical tests
performed in Section \ref{sec:verification}.

Consider a single link $ij$ in a network and assume that $p_i$ and
$p_j$ are given. Then the ODE \eqref{eq:dzdt} for the interface
positions in the link is
\begin{linenomath}
  \begin{align}
    \label{eq:dzdt_p}
    \od{\vec{z}_{ij}}{t} = \frac{q_{ij}\left( \vec{z}_{ij} \right)}{a_{ij}}.
  \end{align}
\end{linenomath}
We further assume that the flow rate in this link is low. This means
that the node and capillary pressures almost balance at the current
interface positions $\vec{z}_{ij}^*$, and thus
$q_{ij}\left(\vec{z}_{ij}^*\right) \approx 0$. Also, we neglect the
dependence of $g_{ij}$ on the interface positions. Now rewrite
\eqref{eq:dzdt_p} in terms $\Delta \vec{z}_{ij} = \vec{z}_{ij} -
\vec{z}_{ij}^*$ and linearize the right hand side around
$\vec{z}_{ij}^*$ to get
\begin{linenomath}
  \begin{align}
    \od{}{t} \Delta \vec{z}_{ij}
    &\approx \frac{q_{ij}\left( \vec{z}_{ij}^* \right)}{a_{ij}}
    + \frac{g_{ij}\left( \vec{z}_{ij}^* \right)}{a_{ij}} \left(
    \sum_{z\in\vec{z}_{ij}^*}\pd{c_{ij}}{z}\right) \Delta \vec{z}_{ij}, \\
    &\approx \frac{g_{ij}\left( \vec{z}_{ij}^*\right)}{a_{ij}} \left(
    \sum_{z\in\vec{z}_{ij}^*}\pd{c_{ij}}{z}\right) \Delta \vec{z}_{ij}, \\
    &= \lambda \Delta \vec{z}_{ij}.
  \end{align}
\end{linenomath}
We can now read off the approximate ODE eigenvalue as
\begin{linenomath}
  \begin{align}
    \lambda = \frac{g_{ij}\left( \vec{z}_{ij}^* \right)}{a_{ij}} \left(
    \sum_{z\in\vec{z}_{ij}^*}\pd{c_{ij}}{z}\right).
  \end{align}
\end{linenomath}
Without loss of generality, we may assume that $\lambda < 0$. If this
is not the case, we interchange the indices $i$ and $j$ and redefine
our spatial coordinate so that $z \to -z$ to get an ODE with negative
$\lambda$. We therefore write the eigenvalue as
\begin{linenomath}
  \begin{align}
    \lambda = -\frac{g_{ij}\left( \vec{z}_{ij}^* \right)}{a_{ij}} \left|
    \sum_{z\in\vec{z}_{ij}^*}\pd{c_{ij}}{z}\right|.
  \end{align}
\end{linenomath}

If the forward Euler method is to be stable on the linearized ODE,
$\lambda \Delta t$ must lie in the stability region of the forward
Euler method \citep{Suli2006},
\begin{linenomath}
  \begin{align}
    \label{eq:fem_stability}
    -2 < \lambda \Delta t < 0.
  \end{align}
\end{linenomath}
This is satisfied if we choose the time step such that
\begin{linenomath}
  \begin{align}
    \label{eq:dt_linear}
    \Delta t < \frac{2 a_{ij}}{g_{ij} \left( \vec{z}_{ij}^* \right)
      \left| \sum_{z\in\vec{z}_{ij}^*}\pd{c_{ij}}{z}\right|}.
  \end{align}
\end{linenomath}
The criterion \eqref{eq:dt_c} is obtained by demanding that
\eqref{eq:dt_linear} be satisfied for all links in the network. If the
advective criterion \eqref{eq:dt_a} is used by itself and the link flow
rates are low, then \eqref{eq:dt_linear} is not necessarily satisfied
for all links and we must expect numerical instabilities from the
forward Euler method.

As the midpoint method has the same real-space stability region as the
forward Euler method \eqref{eq:fem_stability}, the above reasoning and
the criterion \eqref{eq:dt_c} can be applied for the midpoint method
also.

The backward Euler method, on the other hand, is stable if \citep{Suli2006}
\begin{linenomath}
  \begin{align}
    \lambda \Delta t < 0,
  \end{align}
\end{linenomath}
and, because $\lambda$ is negative, it is stable with any positive
$\Delta t$ for this linearized problem.

\section{Jacobian matrix for the semi-implicit method}
\label{sec:jacobian}

In order to solve \eqref{eq:non-linear_system} using the numerical
method described in Section \ref{sec:implementation}, it is necessary
to have the Jacobian matrix of $\vec{F}$. This matrix may be written
as
\begin{linenomath}
  \begin{align}
    \pd{F_i}{p_j^{(n+1)}} &= \delta_{ij} \sum_k \pd{q_{ik}^{(n+1)}}{p_i^{(n+1)}}
    + \left\{ 1 - \delta_{ij} \right\} \pd{q_{ij}^{(n+1)}}{p_j^{(n+1)}}.
  \end{align}
\end{linenomath}

The derivative of $q_{ik}^{(n+1)}$ with respect to $p_i^{(n+1)}$ can
be found by differentiation of \eqref{eq:q_ij_sim} with respect to
$p_i^{(n+1)}$ and application of the chain rule,
\begin{linenomath}
  \begin{align}
    \pd{q_{ij}^{(n+1)}}{p_i^{(n+1)}} 
    &= -g_{ij}^{(n)} + g_{ij}^{(n)} \pd{c_{ij}^{(n+1)}}{p_i^{(n+1)}}, \\
    &= -g_{ij}^{(n)} + g_{ij}^{(n)} \od{c_{ij}^{(n+1)}}{q_{ij}^{(n+1)}}
    \pd{q_{ij}^{(n+1)}}{p_i^{(n+1)}}.
  \end{align}
\end{linenomath}
This can be solved for the desired derivative to yield
\begin{linenomath}
  \begin{align}
    \label{eq:dq_ij_dp_i}
    \pd{q_{ij}^{(n+1)}}{p_i^{(n+1)}} &= -\frac{g_{ij}^{(n)}}{1 -
      g_{ij}^{(n)} \od{c_{ij}^{(n+1)}}{q_{ij}^{(n+1)}}}.
  \end{align}
\end{linenomath}
Herein, the derivative of capillary pressure with respect to flow rate is
\begin{linenomath}
  \begin{align}
    \od{c_{ij}^{(n+1)}}{q_{ij}^{(n+1)}} &= \frac{2\sigma_{\w\n}}{r_{ij}}
    \sum_{z \in \vec{z}_{ij}^{(n+1)}} \left( \pm 1 \right) \sin \left( 2
    \pi \chi_{ij}\left(z\right) \right) \od{\chi_{ij}}{z} \frac{2 \pi
      \Delta t^{(n)}}{ a_{ij}},
  \end{align}
\end{linenomath}
for the specific choice of capillary pressure model given by
\eqref{eq:c_ij}.

As the pore network model is linear in the node pressures, it is
intuitive that the effect on the link flow rate of increasing the
pressure in the node at one end of a link is the same as decreasing
it, by the same amount, in the node at the other end. Thus we may
write
\begin{linenomath}
  \begin{align}
    \label{eq:dq_ij_dp_j_dq_ij_dp_i}
    \pd{q_{ij}^{(n+1)}}{p_j^{(n+1)}} &= - \pd{q_{ij}^{(n+1)}}{p_i^{(n+1)}}.
  \end{align}
\end{linenomath}
This equation may be more formally derived by differentiating
\eqref{eq:q_ij_sim} with respect to $p_j^{(n+1)}$ to get
\begin{linenomath}
  \begin{align}
    \pd{q_{ij}^{(n+1)}}{p_j^{(n+1)}} 
    &= g_{ij}^{(n)} + g_{ij}^{(n)} \pd{c_{ij}^{(n+1)}}{p_j^{(n+1)}}, \\
    &= g_{ij}^{(n)} + g_{ij}^{(n)} \od{c_{ij}^{(n+1)}}{q_{ij}^{(n+1)}}
    \pd{q_{ij}^{(n+1)}}{p_j^{(n+1)}},
  \end{align}
\end{linenomath}
and, solving for the desired derivative,
\begin{linenomath}
  \begin{align}
    \label{eq:dq_ij_dp_j}
    \pd{q_{ij}^{(n+1)}}{p_j^{(n+1)}} 
    &= \frac{g_{ij}^{(n)}}{1 - g_{ij}^{(n)} \od{c_{ij}^{(n+1)}}{q_{ij}^{(n+1)}}}.
  \end{align}
\end{linenomath}
Comparison of \eqref{eq:dq_ij_dp_i} and \eqref{eq:dq_ij_dp_j} gives
the intuitive result \eqref{eq:dq_ij_dp_j_dq_ij_dp_i}.

The addition of $F_m$ \eqref{eq:sim_Q_BC} to the non-linear system for
the specified flow rate boundary condition, introduces some new terms
in the Jacobian matrix of $\vec{F}$. The derivatives of $F_m$ with
respect to the node pressures are
\begin{linenomath}
  \begin{align}
    \pd{F_m}{p_k^{(n+1)}} &= \sum_{ij \in \Omega} \left\{ \delta_{ki}
    \pd{q_{kj}^{(n+1)}}{p_k^{(n+1)}} + \delta_{kj}
    \pd{q_{ik}^{(n+1)}}{p_k^{(n+1)}}\right\},
  \end{align}
\end{linenomath}
where link flow rate derivatives are calculated using
\eqref{eq:dq_ij_dp_i} and \eqref{eq:dq_ij_dp_j_dq_ij_dp_i} and the
derivative with respect to $\Delta P$ is
\begin{linenomath}
  \begin{align}
    \pd{F_m}{\left(\Delta P \right)} &= - \sum_{ij \in \Omega} 
    \frac{g_{ij}^{(n)}}{1 - g_{ij}^{(n)}
      \od{c_{ij}^{(n+1)}}{q_{ij}^{(n+1)}}}.
  \end{align}
\end{linenomath}
The additional terms corresponding to the derivatives with respect to
$\Delta P$ of the mass balance equations for each node $k$ with
unknown pressures are
\begin{linenomath}
  \begin{align}
    \pd{F_k}{\left(\Delta P \right)} &= \sum_{ij \in \Omega} \left\{
    \delta_{kj} - \delta_{ki} \right\} \frac{g_{ij}^{(n)}}{1 -
      g_{ij}^{(n)} \od{c_{ij}^{(n+1)}}{q_{ij}^{(n+1)}}}.
  \end{align}
\end{linenomath}


\end{document}